\documentclass{article}
\usepackage[preprint]{neurips_2026}

\usepackage[utf8]{inputenc}
\usepackage[T1]{fontenc}

\usepackage{url}
\usepackage{microtype}
\usepackage{xcolor}

\usepackage{amsmath}
\usepackage{amssymb}
\usepackage{amsfonts}
\usepackage{mathtools}
\usepackage{amsthm}
\usepackage{nicefrac}

\usepackage{graphicx}
\usepackage{booktabs}
\usepackage{multirow}
\usepackage{arydshln}
\usepackage{caption}
\usepackage{subcaption}

\usepackage{algorithm}
\usepackage{algpseudocode}

\usepackage{listings}
\usepackage[most]{tcolorbox}
\tcbuselibrary{listings,skins,breakable}

\usepackage{hyperref}
\usepackage[capitalize,noabbrev]{cleveref}

\lstdefinestyle{compactprolog}{
  basicstyle=\ttfamily\scriptsize,
  columns=fullflexible,
  keepspaces=true,
  showstringspaces=false,
  breaklines=true,
  breakatwhitespace=false,
  aboveskip=0pt,
  belowskip=0pt
}

\lstdefinestyle{appendixprolog}{
  basicstyle=\ttfamily\footnotesize,
  columns=fullflexible,
  keepspaces=true,
  showstringspaces=false,
  breaklines=true,
  breakatwhitespace=false,
  aboveskip=0pt,
  belowskip=0pt
}

\newtcolorbox{compactalgobox}[1][]{%
  enhanced,
  sharp corners,
  colback=white,
  colframe=black,
  boxrule=0.45pt,
  boxsep=0pt,
  left=1mm,
  right=1mm,
  top=0.8mm,
  bottom=0.8mm,
  before skip=0pt,
  after skip=2pt,
  #1
}

\newtcblisting{compactcodebox}[1][]{%
  listing only,
  enhanced,
  sharp corners,
  colback=white,
  colframe=black,
  boxrule=0.45pt,
  boxsep=0pt,
  left=1mm,
  right=1mm,
  top=0.8mm,
  bottom=0.8mm,
  before skip=0pt,
  after skip=2pt,
  listing options={style=compactprolog},
  #1
}

\newtcblisting{breakablecodebox}[1][]{%
  listing only,
  breakable,
  enhanced,
  sharp corners,
  colback=white,
  colframe=black,
  boxrule=0.45pt,
  boxsep=0pt,
  left=1.5mm,
  right=1.5mm,
  top=1mm,
  bottom=1mm,
  before skip=2pt,
  after skip=2pt,
  listing options={style=appendixprolog},
  #1
}

\lstdefinestyle{prologbox}{
  basicstyle=\ttfamily\footnotesize,
  columns=fullflexible,
  keepspaces=true,
  showstringspaces=false,
  breaklines=true,
  breakatwhitespace=false,
  frame=single,
  framerule=0.4pt,
  framesep=4pt,
  xleftmargin=0pt,
  xrightmargin=0pt,
  aboveskip=2pt,
  belowskip=2pt
}

\newtcolorbox{longprompt}[2][]{%
  listing only,
  breakable,
  enhanced,
  sharp corners,
  colback=gray!5!white,
  colframe=gray!75!black,
  title=\textbf{#2},
  fonttitle=\sffamily,
  fontupper=\ttfamily\small,
  listing options={
    basicstyle=\ttfamily\small,
    breaklines=true,
    breakatwhitespace=false,
    columns=fullflexible,
    keepspaces=true,
    showstringspaces=false
  },
  boxrule=0.6pt,
  arc=1pt,
  left=5pt,
  right=5pt,
  top=5pt,
  bottom=5pt,
  #1
}

\theoremstyle{plain}

\theoremstyle{definition}

\theoremstyle{remark}

\title{Procedural Refinement by LLM-driven Algorithmic Debugging for ARC-AGI-2}

\author{%
  Yu-Ning Qiu\textsuperscript{*} \quad
  Lin-Feng Zou\textsuperscript{*} \quad
  Jiong-Da Wang \\
  Xue-Rong Yuan \quad
  Wang-Zhou Dai\textsuperscript{$\dagger$} \\
  Nanjing University \\
  Nanjing, China \\
  \texttt{dai.wzero@gmail.com} \\
  \\[-0.2em]
  {\small \textsuperscript{*}Equal contribution. \quad
  \textsuperscript{$\dagger$}Corresponding author.}
}

\begin{document}

\maketitle

\begin{abstract}
In high-complexity abstract reasoning, a system must infer a latent rule from a
few examples or structured observations and apply it to unseen instances.  LLMs
can express such rules as programs, but ordinary conversation-based refinement
is largely outcome-level: it observes that an answer or output is wrong without
formally re-checking which abstraction, relation, or transformation justified
that outcome.  We propose \emph{Abduction-Based Procedural Refinement} (ABPR), a
neuro-symbolic refinement approach that couples an LLM with a Prolog
meta-interpreter.  ABPR treats each candidate program as an executable
declarative hypothesis of the latent rule and reifies its SLD goal--subgoal
resolution into compact proof-tree-style derivations, following Shapiro's
algorithmic program debugging (APD).  In this view, refinement is not merely
code-level debugging, but semantic re-checking of the model's hypothesised rule.
We evaluate ABPR primarily on ARC-AGI-2, a challenging few-shot abstract rule
induction benchmark over grid transformations.  ABPR with Gemini-3-Flash
achieves 56.67\% Pass@2, while GPT-5.5 xHigh with ABPR reaches 98.33\% Pass@2
on the public evaluation set.  Supplementary experiments on fill-in-the-blank
I-RAVEN-X and A-I-RAVEN adaptations provide evidence that the same
trace-guided framework extends beyond ARC-specific grid tasks to RAVEN-style
relational and analogical abstraction.  Repeated-run and sensitivity analyses show that parallel trace-guided search
reduces stochastic variance as search breadth and total search depth increase.
\end{abstract}

\section{Introduction}

Although large language models (LLMs) have made remarkable progress in
generating outputs that appear logically coherent and contextually plausible,
they do not provide guarantees of correctness or formal validity in their
reasoning processes, particularly on algorithmically structured
tasks~\cite{rossi2025aaai}.  This limitation is critical when LLMs are expected
to operate on complex tasks with limited supervision, especially in settings
where the task abstractions and problem-solving patterns are not covered in the
training data~\cite{brown2020language}.  Hence, the ability to recover from
erroneous outputs, often referred to as ``self-correction'', becomes a central
requirement for LLM-based systems~\citep{kamoi2024can}.

However, effective correction remains a step-by-step process that requires
structured search, planning, and diagnostic evidence, rather than a single
conversational revision.  A growing body of recent work has shown that LLMs
struggle to perform such correction robustly when relying solely on intrinsic or
conversational feedback; in some cases, post-revision outputs are even worse
than first-pass responses~\cite{huang2023large,kamoi2024can,chen2025studying}.
Even when augmented with chain-of-thought prompting and external compilers or
program executors, evidence shows that iterative LLM-based code repair can
degrade across successive attempts~\cite{adnan2025measuring}.

Tool-augmented correction partially addresses this problem by exposing
verifiable signals such as test results, compiler messages, solver outcomes, or
program executions.  These signals are valuable because they ground revision in
external evidence rather than in the model's own conversational judgement.
However, for abstract rule induction, the object to be refined is the
hypothesised rule itself.  A candidate program may execute exactly as written
while encoding the wrong abstraction.  For example, a reachability hypothesis
may define \texttt{reachable} as a two-hop relation rather than as transitive
closure.  A failed query or breakpoint trace can reveal a concrete failed check,
but the useful diagnosis is semantic: the induced relation is under-defined and
should recurse on \texttt{reachable}.  This distinction, detailed in
Appendix~\ref{app:semantic_example}, motivates a refinement mechanism that
inspects formal derivations of candidate rules, rather than relying only on
final outcomes or execution states.

This limitation motivates revisiting a classical symbolic-AI foundation for
semantic diagnosis: Shapiro's framework of \emph{Algorithmic Program Debugging}
(APD)~\cite{shapiro1982algorithmic}.  APD formalizes debugging as a structured
process of inspecting derivations and identifying which program judgement is
inconsistent with the intended semantics, rather than performing an ad hoc
sequence of revisions.  For abstract rule induction, the generated program is
an explicit hypothesis of a latent rule, so refinement should re-check the
semantic commitments made by that hypothesis.  Prolog provides a direct
operational realization of this idea: SLD resolution decomposes a query into
goals and subgoals, allowing a meta-interpreter to reify the reasoning process
into compact proof-tree-style derivations that provide explicit semantic
evidence for localized refinement.

\begin{figure}
    \centering
    \includegraphics[width=\linewidth]{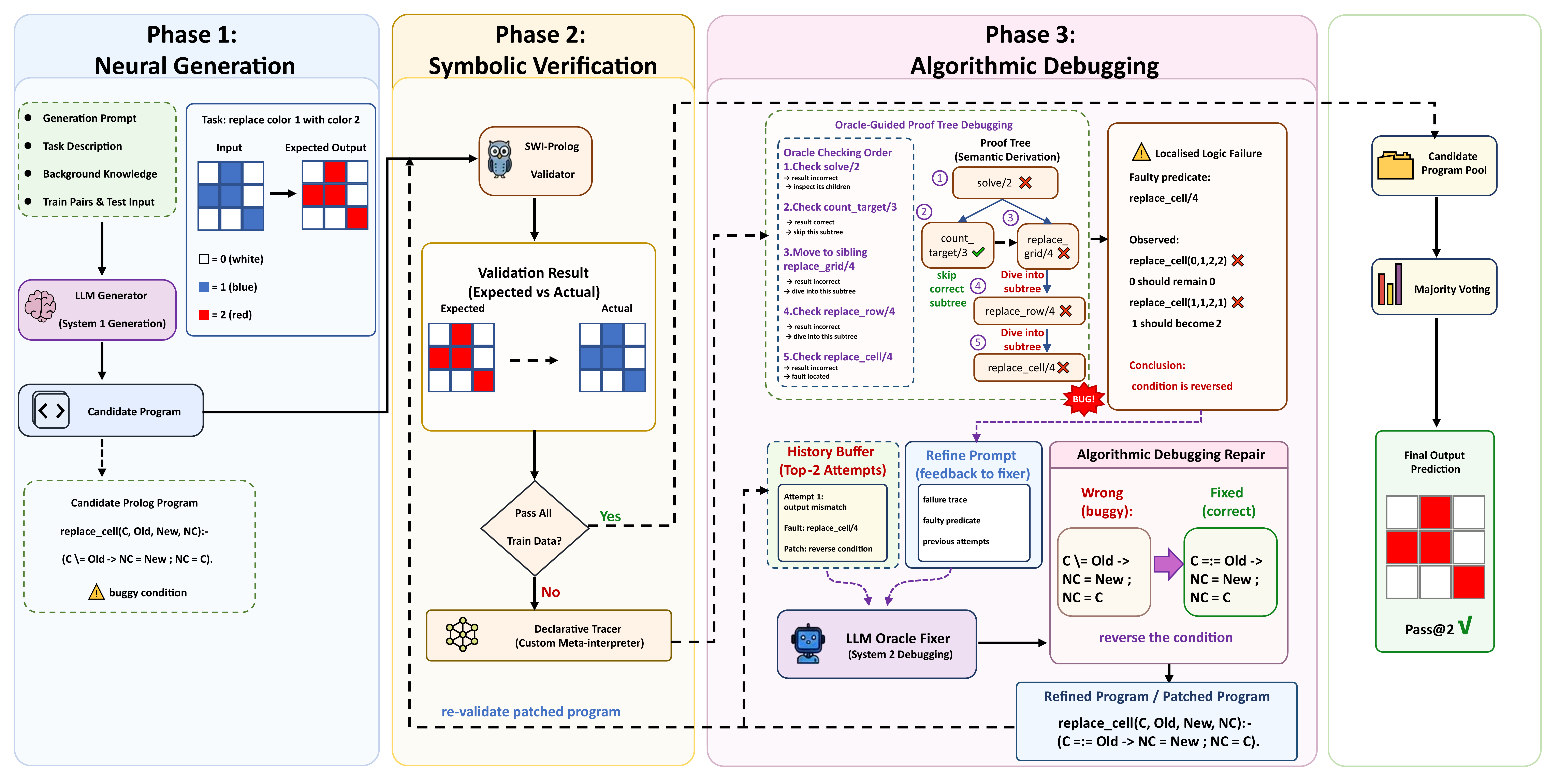}
    \captionof{figure}{Overview of ABPR: neural generation, symbolic verification, and semantic-trace-guided procedural refinement.}
    \label{fig:overview}
\end{figure}

In this paper, we formulate \emph{procedural refinement} as the problem of
revising executable hypotheses for abstract rule induction.  Rather than asking
an LLM to conversationally revise a final answer, we treat each candidate
program as a declarative hypothesis about the latent rule and refine it through
a structured, multi-step process guided by abductive
reasoning~\citep{kakas1992abductive,zhou2019abductive}.  Based on APD, we
implement \emph{Abduction-Based Procedural Refinement} (ABPR), a neuro-symbolic
approach that couples an LLM with a Prolog meta-interpreter.  The
meta-interpreter reifies goal--subgoal resolution into compact tree-structured
semantic derivations, allowing ABPR to localize and revise faulty reasoning
steps in the hypothesised rule, as illustrated in
Figure~\ref{fig:overview}.

We evaluate ABPR primarily on ARC-AGI-2~\cite{chollet2025arc}, a challenging
few-shot abstract rule induction benchmark where success requires discovering
and applying latent grid-transformation rules from only a handful of examples.
This setting is well suited for studying procedural refinement: failures often
come from incorrect intermediate abstractions or transformation rules, while
outcome-level revision provides little guidance about which part of the
hypothesised rule should be revised.  ABPR with Gemini-3-Flash achieves
56.67\% Pass@2 on ARC-AGI-2, while GPT-5.5 xHigh with ABPR reaches 98.33\%
Pass@2 on the public evaluation set.  Ablations show that these gains are
driven primarily by trace-guided semantic refinement rather than by execution
feedback or conversational revision.

To test whether this framework is tied to ARC-specific grid transformations, we
also evaluate ABPR on fill-in-the-blank adaptations of I-RAVEN-X~\citep{camposampiero2025iravenx} and
A-I-RAVEN~\citep{malkinski2025airaven}.  These supplementary non-ARC experiments target RAVEN-style
relational and analogical abstraction, and provide evidence that the same
semantic-trace-guided refinement mechanism extends beyond the ARC setting.

\section{Preliminaries}
\label{sec:pre}

\subsection{Abstract rule induction}
\label{subsec:abstract_rule_induction}

We study \emph{abstract rule induction}: a task setting in which a system must
infer a latent rule from a small number of examples or structured observations
and apply that rule to unseen instances.  In ARC-AGI-2, the latent rule is a
grid transformation inferred from input--output examples; in RAVEN-style matrix
reasoning, it is a relational or analogical rule over panel attributes.  In
both cases, the central difficulty is not merely producing a plausible final
answer, but inducing a rule-level hypothesis whose intermediate abstractions,
relations, and transformations remain valid under new cases.

In this paper, we represent such hypotheses as executable declarative programs.
Let $\mathcal{B}$ denote a domain background theory, such as grid operations,
object relations, or panel-attribute predicates.  A candidate program $P$
specifies a hypothesised rule over $\mathcal{B}$.  Given training observations
$\mathcal{E}_{\mathrm{train}}$, the desired condition is that
$\mathcal{B}\cup P$ explains the observed examples, written abstractly as
$\mathcal{B}\cup P \models \mathcal{E}_{\mathrm{train}}$.  When this condition
fails, the error is not necessarily a software defect in the usual sense; it may
be an error in the induced abstraction, relation, or transformation rule.  This
is the setting in which we define procedural refinement.

\begin{figure}
    \centering
    \includegraphics[width=\linewidth]{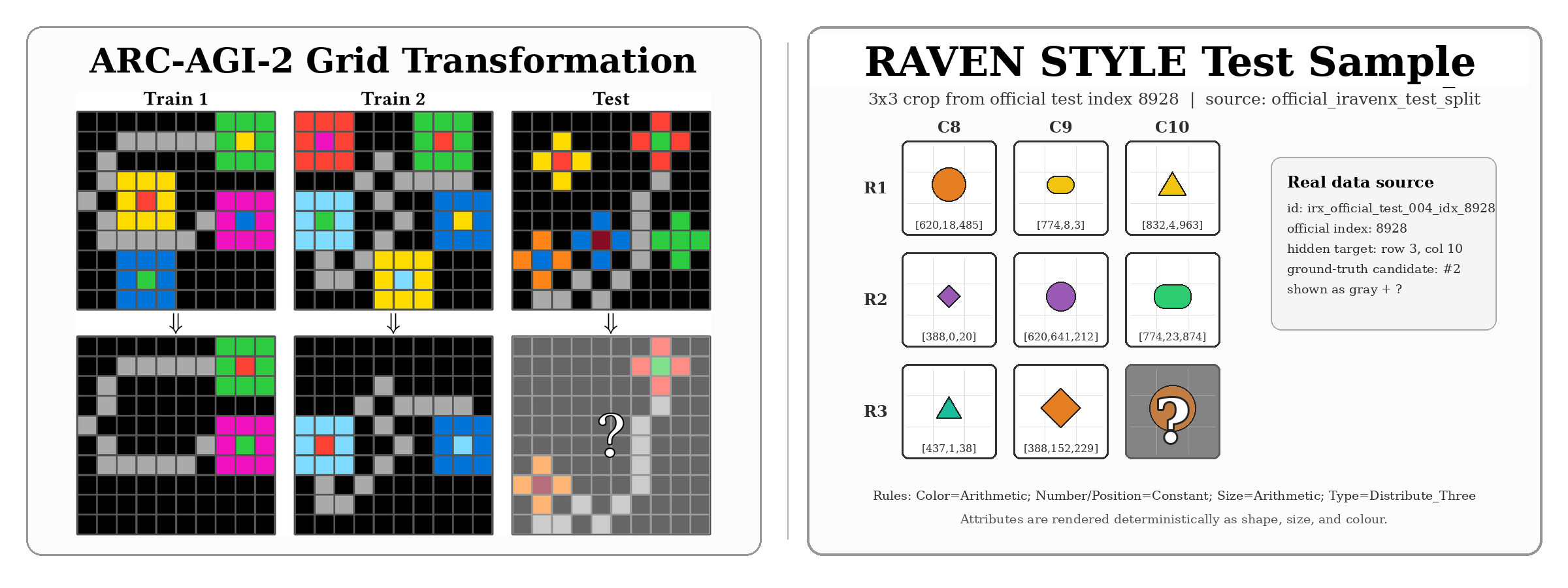}
    \captionof{figure}{Representative ARC-AGI-2 grid transformation and
    supplementary RAVEN-style matrix reasoning tasks.}
    \label{fig:problem_examples}
\end{figure}

\subsection{Algorithmic Program Debugging}
\label{subsec:apd}

Algorithmic Program Debugging (APD), introduced by
Shapiro~\citep{shapiro1982algorithmic}, provides a formal foundation for
semantic diagnosis in logic programs.  Rather than treating debugging as an
ad hoc inspection of variables or control flow, APD inspects the logical
structure of a derivation and asks which local judgement is inconsistent with
the intended semantics.

The central object in APD is a \emph{debugging tree}, or proof tree.  Its root
corresponds to the initial query, internal nodes correspond to subgoals or
predicate invocations, and leaves correspond to primitive facts or built-in
relations.  Traditionally, APD traverses this tree and queries an oracle about
the semantic validity of selected nodes.  In our setting, the oracle role is
played by an LLM, but the object being inspected is still a formal derivation
rather than a free-form conversation.

Formally, let $P$ be a logic program and let $M$ be the intended interpretation,
i.e., the set of ground atoms that should be considered true under the target
rule.  For an incorrect successful derivation, a clause instance
$H \leftarrow B_1,\ldots,B_k$ is locally faulty when the body literals are
semantically valid, $B_i\in M$ for all $i$, but the head is not, $H\notin M$.
This localizes the error to a single inference step: the sub-computations are
accepted, but the conclusion they support is not.  For missing expected
answers, the complementary diagnosis concerns a required judgement in $M$ for
which the program fails to construct a valid derivation.  Thus APD turns global
mismatches into localized semantic questions about the candidate rule.

\subsection{Procedural refinement as abduction}
\label{subsec:PR}

We formalize \emph{procedural refinement} as abductive revision of an
executable hypothesis.  Let $P_0$ be the initial program generated by an LLM,
$\mathcal{B}$ the background theory, and $\mathcal{E}_{\mathrm{train}}$ the
training observations.  A refinement operator $\rho$ maps a program to locally
revised candidates, and its closure is
\[
\rho^*(P_0)=\rho^0(P_0)\cup\rho^1(P_0)\cup\cdots,\qquad
\rho^0(P_0)=\{P_0\}.
\]
The refinement objective is to find a program
\[
P^* \in \rho^*(P_0)
\quad\text{such that}\quad
\mathcal{B}\cup P^* \models \mathcal{E}_{\mathrm{train}},
\]
subject to the constraint that revisions are guided by localized semantic
evidence rather than unconstrained whole-program rewriting.

At iteration $i$, APD constructs a derivation tree for the current program
$P_i$ and identifies a small set of suspicious semantic judgements.  The LLM
then abduces a local correction $\Delta_i$ that would explain and repair the
observed discrepancy:
\[
P_{i+1}=P_i\oplus \Delta_i.
\]
Here $\Delta_i$ may add, delete, or modify a predicate condition, recursive
subgoal, or relation binding.  This view makes refinement an abductive
inference problem~\citep{kakas1992abductive,zhou2019abductive}: find a minimal
change to the hypothesised rule that reconciles its derivation with the
training observations.

\subsection{Logic programming and meta-interpreters}
\label{subsec:logic_programming}

We instantiate ABPR in Prolog~\citep{lloyd2012foundations} because its
operational semantics provide a direct realization of the derivation structures
required by APD.  Prolog evaluates queries by SLD resolution, repeatedly
reducing goals into subgoals through unification and clause
selection~\citep{robinson1965machine}.  As a result, a candidate rule is not
only executable; its reasoning process can also be represented as a structured
goal--subgoal derivation.

We use a Prolog \emph{meta-interpreter} to reify this process.  Rather than
recording a raw execution log, the meta-interpreter constructs a compact
tree-structured derivation whose nodes include the current goal, relevant
variable bindings, the clause instance being applied, success or failure
status, and child subgoals~\citep{o2009craft}.  These proof-tree-style
derivations provide the semantic evidence used by ABPR to localize faulty
reasoning steps and propose targeted revisions.  Other proof-producing
formalisms may support analogous traces, but Prolog makes this object directly
available through SLD resolution and meta-interpretation.

\subsection{The ARC-AGI-2 benchmark}
\label{subsec:arc}

ARC-AGI-2~\citep{chollet2025arc} is our primary evaluation setting.  Each task
contains a small number of input--output grid examples and one or more test
inputs.  Solving a task requires inducing the latent transformation rule from
the examples and applying it to the unseen test input.  This makes ARC-AGI-2 a
natural test bed for procedural refinement: first-pass hypotheses often fail
because they capture the wrong object abstraction, spatial relation, or
transformation condition, while outcome-level feedback gives little indication
of which part of the hypothesised rule should be revised.

\subsection{RAVEN-style benchmarks}
\label{subsec:raven}
We additionally use fill-in-the-blank adaptations of I-RAVEN-X~\citep{camposampiero2025iravenx} and A-I-RAVEN~\citep{malkinski2025airaven} as
supplementary non-ARC evaluations.  These settings test whether the same
semantic-trace-guided refinement mechanism extends from grid transformations to
RAVEN-style relational and analogical abstraction, providing evidence that ABPR is
not tied to ARC-specific grid engineering.

\section{Abduction-Based Procedural Refinement}
\label{sec:method}
This section introduces \emph{Abduction-Based Procedural Refinement} (ABPR),
illustrated in Figure~\ref{fig:overview}.  ABPR gives a formal account of
self-refinement for executable abstract-rule hypotheses by decomposing the
global search induced by APD into a sequence of localized abductive subproblems.
By representing candidate rules as declarative programs and reifying their
goal--subgoal derivations with a symbolic meta-interpreter, ABPR bridges
neural generative flexibility with semantic verification, yielding a refinement
process that is localized and auditable.


\subsection{Framework}
\label{subsec:formal_framework}


As introduced in Section~\ref{subsec:PR}, the objective of ABPR is to find an
optimal hypothesis $H^*$ within the \textbf{refinement closure}
$\rho^*(H_{\text{init}})$ of the initial program $H_{\text{init}}$, such that:
\begin{align}
\mathcal{B}\cup H^*\models \mathcal{E}_{\text{train}}, 
\end{align}
where $\rho$ is the \emph{refinement operator}. Formally, $\rho$ is a mapping that defines the set of all admissible modifications for a given program. The search space is the transitive closure defined inductively as $\rho^*(H) = \bigcup_{i=0}^{\infty} \rho^i(H)$ with $\rho^0(H) = \{H\}$.

\noindent\textbf{Hypothesis initialisation.}
Inspired by \emph{Inverse Entailment} \citep{muggleton1995inverse}, We derive
the starting state $H_{\text{init}}$ by adopting a strategy.
Specifically, given the background knowledge $\mathcal{B}$ and the examples
$\mathcal{E}$, we construct  \emph{bottom clause}  as the most specific logical
explanation to the training data. In the ARC-AGI-2 task, $\mathcal{B}$ contains
the input matrices and some pre-defined primitive predicates for parsing them,
while $\mathcal{E}$ contains the corresponding output matrices.

Then, different to the traditional Inductive Logic Programming (ILP) methods, the
LLM-driven ABPR prompts an LLM to generalise the bottom clauses into a (possibly
imperfect) logic program to form $H_{\text{init}}$.


\noindent\textbf{Refinement via stochastic search.}
Theoretically, the refinement operator $\rho(H)$ induces a \emph{hypothesis
  space} of candidate programs. Traditional symbolic learning methods define a
set of refinement operators that modify an initial hypothesis $H$---for example,
by adding or deleting rules or literals---and employ a scoring function to guide
the search toward improved refinements. Under this paradigm, classical APD is a
deterministic procedure: assuming access to an oracle with ground-truth
knowledge, it performs formal abductive reasoning over the derivation tree to
identify the \emph{minimal inconsistent subtree}.

In our framework, both the oracle and the abductive reasoner are approximated by
Large Language Models (LLMs), motivated by the
\emph{discriminator--generator
  gap}~\cite{cobbe2021training,saunders2022self,lightman2023lets}, which shows
that verifying local reasoning steps is substantially easier than synthesising a
complete reasoning trajectory.
ABPR leverages this idea by reducing program synthesis/refinement to a sequence
of low-entropy, verifiable sub-problems.
Unlike intrinsic self-correction methods, which suffer from
circularity~\cite{huang2023large}, our approach relies on semantic derivation
traces, aligning with advances observed in recent tool-augmented critique
frameworks~\cite{gou2024critic,ridnik2024alphacodium}.

However, because LLM generation is
fundamentally a process of sampling from a learned conditional probability
distribution over the vocabulary given an input
context~\cite{holtzman2019curious}, integrating LLMs into the refinement loop
introduces inherent stochasticity.

As a result, the originally deterministic diagnosis trajectory becomes
probabilistic. A single refinement trajectory of our neuro-symbolic APD can
therefore be formalized as sampling a successor hypothesis from a complex
conditional distribution induced by the composition of the neural oracle and the
LLM-based repairer:
\[
H_{t+1} \sim P_{\text{APD}}(\cdot \mid H_t, \tau,\mathcal{E}_{\text{train}}, \mathcal{B}), \quad \text{where } H_{t+1} \in \rho^*(H_t).
\]
In practice, rather than seeking a single ``correct'' fix, the procedure
stochastically explores high-likelihood regions of the refinement closure,
guided by the joint confidence of the neural components.



\subsection{Declarative trace generation}
\label{subsec:traces}

To enable neuro-symbolic APD to diagnose and revise faulty reasoning steps, the
system must recover the semantic structure underlying a candidate rule's
derivation.  We therefore represent the derivation signal $\tau$ as a
\textbf{proof-tree-style semantic derivation}, adapting the classical APD
formalism to link each inferred outcome with the clauses and subgoals that
support it.

We extract these traces using a lightweight Prolog meta-interpreter implemented
with SWI-Prolog~\cite{WielemakerSTL12:SWI}; Appendix~\ref{app:trace_meta_interpreter}
shows the core implementation. The tracer extends standard SLD resolution by
recursively constructing terms of the form
\texttt{node(Type, InputSnapshot, OutputSnapshot, Children)}, where each node
records the invoked predicate, its input/output bindings, and the selected
sub-derivations.

This construction captures logical dependencies and variable bindings at each
inference step, yielding a structured semantic proof tree. The resulting
derivation tree serves as evidence for fault localisation and subsequent
hypothesis refinement.

\begin{algorithm}[t]
\caption{ABPR: iterative procedural refinement.}
\label{alg:abpr}
\begin{algorithmic}[1]
\Require Background knowledge $\mathcal{B}$; training examples
$\mathcal{E}_{\mathrm{train}}$; total search depth
$T_{\mathrm{depth}}\ge 1$; history size $K=2$
\Ensure Refined program hypothesis $\widehat{H}$

\State $H_0 \gets \mathrm{Initialize}(\mathcal{B}, \mathcal{E}_{\mathrm{train}})$
\State $\mathbb{H} \gets \{H_0\}$

\For{$t = 1,\ldots,T_{\mathrm{depth}}-1$}
    \State $H_b \gets \mathrm{Latest}(\mathbb{H})$
    \State $\tau \gets \mathrm{Trace}(H_b, \mathcal{E}_{\mathrm{train}})$
    \State $H_n \sim P_{\mathrm{APD}}(\cdot \mid H_b, \tau, \mathcal{E}_{\mathrm{train}}, \mathcal{B})$

    \If{$\mathrm{IsConsistent}(H_n, \mathcal{E}_{\mathrm{train}})$}
        \State \Return $H_n$
    \EndIf

    \State $s(H) \gets \mathrm{Coverage}(H, \mathcal{E}_{\mathrm{train}})$
    \State $\mathbb{H} \gets \mathrm{TopK}(\mathbb{H} \cup \{H_n\}, K, s)$
\EndFor

\State \Return $\operatorname*{arg\,max}_{H \in \mathbb{H}} \mathrm{Coverage}(H, \mathcal{E}_{\mathrm{train}})$
\end{algorithmic}
\end{algorithm}

\subsection{The neuro-symbolic refinement procedure}
\label{subsec:locate_repair}
ABPR proceeds as a state machine over executable hypotheses. Each transition
corresponds to one iteration of the procedural refinement cycle, updating the
current hypothesis from $H_t$ to $H_{t+1}$.

\noindent\textbf{Step 1: Fault localisation.}
Given a trace $\tau$ produced by a failed hypothesis $H_t$, the LLM Oracle
performs a top-down diagnostic search over the computation tree to identify a
set of \emph{candidate faulty nodes}, denoted by $\mathcal{N}^*$. Conceptually,
this step serves as a \textbf{semantic pruning mechanism}: by isolating
$\mathcal{N}^*$, we condition the distribution $P_{\text{APD}}$ on the observed
logical discrepancies, thereby restricting the subsequent search to a focused
subspace of the refinement closure $\rho^*(H_t)$ rather than the full program
space.

\noindent\textbf{Step 2: Abductive refinement.}
Conditioned on the identified faulty set $\mathcal{N}^*$, the refinement
operator performs abductive hypothesis revision to construct the successor
state. Specifically, it synthesises a modified hypothesis $H_{t+1}$ that
resolves the local inconsistencies associated with $\mathcal{N}^*$ while
preserving the global structure of the program.

\noindent\textbf{Remarks on the implementation.}
In practice, these steps are implemented as a fixed verifier-in-the-loop
refinement procedure controlled by prompts. The LLM is instructed to (i) carry
out APD-style inspection to localise suspicious clause instances and (ii)
generate targeted revisions for the clauses associated with $\mathcal{N}^*$.
Taken together, this module functions as a stochastic sampler from the
conditional distribution $P_{\text{APD}}(\cdot \mid H_t,
\mathcal{E}_{\text{train}},\mathcal{B})$, implementing semantic state
transitions in the hypothesis space by coupling abductive reasoning with formal
code generation.

Algorithm~\ref{alg:abpr} shows the pseudocode of ABPR. It initialises the search
using bottom clauses and then iteratively explores the refinement hypothesis
space, with the neuro-symbolic APD acting as a stochastic transition operator
between successive hypotheses (Line~6). The symbolic characterisation of
\emph{procedural refinement} and the APD framework enables a natural realisation
of a \textbf{best-first search} strategy. To exploit this structure, the
algorithm computes coverage scores and updates a bounded history buffer
$\mathbb{H}$ via a top-$k$ selection mechanism (Lines~10--11), prioritising
high-coverage candidate hypotheses for further refinement in subsequent iterations.

As ABPR performs iterative exploration of a large hypothesis space, individual
refinement trajectories are sensitive to initial conditions and may converge
prematurely to suboptimal solutions.
To improve robustness, we adopt an ensemble-based strategy with
diversity-prioritised aggregation following prior art~\cite{Poetiq2025Frontier}.
Specifically, we perform a \emph{parallel search} in the refinement 
hypothesis space by initiating multiple ABPR instances.

\section{Experiments}
\label{sec:experiments}

We evaluate ABPR in ARC-AGI-2 (public evaluation dataset) to address
the following research questions:
\begin{itemize}
  \item \textbf{RQ1:} Does ABPR improve end-to-end accuracy on ARC-AGI-2?
  \item \textbf{RQ2:} How much do semantic derivation traces contribute to
    performance?
  \item \textbf{RQ3:} How sensitive is ABPR to the quality of the initial hypothesis?
  \item \textbf{RQ4:} Does ABPR generalise beyond ARC-style grid transformation tasks?
\end{itemize}



\subsection{Settings}
Each ARC-AGI-2 task consists 2--5 training input--output pairs and 1--3 test pairs.
Following the official ARC-AGI evaluation protocol, we measure performance using
the standard \textbf{Pass@2} metric~\cite{chen2021evaluating}, under which a task
is considered solved if the correct output is produced within two allowed submissions.




In our experiments, we set the number of parallel threads $N=8$
following~\cite{Poetiq2025Frontier}, where each ABPR thread $A_i$ is initialised
with a distinct random seed $\xi_i$, which determines both the ordering of training
examples and the stochastic choices made within the LLM-based refinement
operator~\cite{dietterich2000ensemble}.
We set the sampling temperature to $1.0$ and intentionally randomise the seed of
each thread to increase the diversity of first-pass program hypotheses. This
choice broadens the initial hypothesis distribution rather than making each
single trajectory deterministic. The resulting variance is controlled by sufficient sampling over the ensemble
size $N$ and total search depth $T_{\mathrm{depth}}$. Here
$T_{\mathrm{depth}}=1$ denotes the first-pass generation only, while
$T_{\mathrm{depth}}>1$ includes $T_{\mathrm{depth}}-1$ subsequent
verifier-guided repair rounds. As $N$ and $T_{\mathrm{depth}}$ increase,
run-to-run variation empirically contracts.
Appendix~\ref{app:variance_reproducibility} reports the repeated-run variance and
heatmap analysis supporting this convergence, and documents the exact data
sources used for reproducibility.


Then, each process $A_i$ runs ABPR to a total search depth of
$T_{\mathrm{depth}}=10$, corresponding to one initial generation followed by up
to nine verifier-guided repair rounds. This produces a final program hypothesis
$H_i$  together with its execution outcomes on the
training set, among which a diversity-prioritised voting is performed to select
the top-$2$ programs that maximise both consensus size
and verification consistency~\cite{kuncheva2003measures}.

\paragraph{Feedback conditions and baselines.}
Our experiments compare feedback conditions under the same verifier and task
protocol.  \emph{Direct} generation receives no iterative feedback.
\emph{Execution-feedback repair} receives failed outputs, verifier outcomes,
and runtime errors, and can revise the program iteratively, but does not receive
APD proof-tree derivations.  \emph{ABPR repair} uses the same verifier feedback
and additionally receives semantic derivation traces for localized refinement.
Python direct and Python execution-feedback rows are reported only in the
supplementary RAVEN-style setting, where the symbolic panel interface admits a
compact no-trace Python reference under the same interactive verifier; ARC
ablations are kept within a fixed Prolog hypothesis space to isolate the effect
of semantic derivation traces.


\begin{table}[t]
\centering
\caption{ARC-AGI-2 Pass@2 results under matched within-block first-pass and refinement settings.
The table separates the model that produces the initial program from the model that performs repair.
Cross-model rows report the initialization pool used for the corresponding cross-model repair experiment.}
\label{tab:consolidated_abpr}
\scriptsize
\setlength{\tabcolsep}{2.5pt}
\setlength{\dashlinedash}{1.4pt}
\setlength{\dashlinegap}{2.8pt}
\newcommand{\modelsep}{\addlinespace[1.6pt]\cdashline{1-6}\addlinespace[1.6pt]}
\newcommand{\blocksep}{\specialrule{1.40pt}{1.2pt}{1.2pt}}
\begin{tabular*}{\linewidth}{@{\extracolsep{\fill}}lllccc@{}}
\toprule
\textbf{Experiment setting} & \textbf{Initial model} & \textbf{Repair model} & \textbf{Iter. repair} & \textbf{ABPR trace} & \textbf{Pass@2} \\
\blocksep
\multicolumn{6}{l}{\textit{Full same-model comparison: first pass $\rightarrow$ execution-feedback repair $\rightarrow$ ABPR repair}} \\
\quad Gemini-3-Flash: first pass & Gemini-3-Flash & -- & $-$ & $-$ & 34.03\% \\
\quad Gemini-3-Flash: execution-feedback repair & Gemini-3-Flash & Gemini-3-Flash & $+$ & $-$ & 49.17\% \\
\quad Gemini-3-Flash: ABPR repair & Gemini-3-Flash & Gemini-3-Flash & $+$ & $+$ & \textbf{56.67\%} \\
\modelsep
\quad Gemini-3-Pro: first pass & Gemini-3-Pro & -- & $-$ & $-$ & 31.11\% \\
\quad Gemini-3-Pro: execution-feedback repair & Gemini-3-Pro & Gemini-3-Pro & $+$ & $-$ & 43.33\% \\
\quad Gemini-3-Pro: ABPR repair & Gemini-3-Pro & Gemini-3-Pro & $+$ & $+$ & 53.33\% \\
\blocksep
\multicolumn{6}{l}{\textit{Across-capability backbones: first pass $\rightarrow$ ABPR repair}} \\
\quad GPT-5.2 low: first pass & GPT-5.2 low-reasoning & -- & $-$ & $-$ & 8.33\% \\
\quad GPT-5.2 low: ABPR repair & GPT-5.2 low-reasoning & GPT-5.2 low-reasoning & $+$ & $+$ & 30.00\% \\
\modelsep
\quad GPT-5.5 xHigh: first pass & GPT-5.5 xHigh & -- & $-$ & $-$ & 85.00\% \\
\quad GPT-5.5 xHigh: ABPR repair & GPT-5.5 xHigh & GPT-5.5 xHigh & $+$ & $+$ & \textbf{98.33\%} \\
\modelsep
\quad Claude-Sonnet-4.5: first pass & Claude-Sonnet-4.5 & -- & $-$ & $-$ & 3.75\% \\
\quad Claude-Sonnet-4.5: ABPR repair & Claude-Sonnet-4.5 & Claude-Sonnet-4.5 & $+$ & $+$ & 5.00\% \\
\modelsep
\quad Qwen3-Max: first pass & Qwen3-Max & -- & $-$ & $-$ & 0.83\% \\
\quad Qwen3-Max: ABPR repair & Qwen3-Max & Qwen3-Max & $+$ & $+$ & 2.50\% \\
\blocksep
\multicolumn{6}{l}{\textit{Cross-model initialization/refinement}} \\
\quad Claude-Sonnet-4.5: initial program only & Claude-Sonnet-4.5 & -- & $-$ & $-$ & 3.75\% \\
\quad Claude-Sonnet-4.5 init + Claude ABPR & Claude-Sonnet-4.5 & Claude-Sonnet-4.5 & $+$ & $+$ & 5.00\% \\
\modelsep
\quad Gemini-3-Pro: initial program only & Gemini-3-Pro & -- & $-$ & $-$ & 27.92\% \\
\quad Gemini-3-Pro init + Claude ABPR & Gemini-3-Pro & Claude-Sonnet-4.5 & $+$ & $+$ & 42.92\% \\
\bottomrule
\end{tabular*}

\vspace{2pt}
\begin{minipage}{\linewidth}
\footnotesize
\emph{Note.} Cross-model initialization-only rows report the initialization
pool used for the corresponding cross-model repair run. The Gemini-3-Pro row is
from an independent stochastic sweep at sampling temperature
$T_{\mathrm{sample}}=1.0$, whereas the Claude row reuses the first-pass programs
from the Claude same-model run.
\end{minipage}
\end{table}

\subsection{RQ1: Overall performance}
To assess whether ABPR improves executable abstract rule induction, we
evaluate its performance across a range of LLMs with markedly different reasoning
profiles.
Specifically, we consider the \textbf{Gemini-3 series} (Flash and Pro
Preview)~\cite{Gemini3TechReport2025}, \textbf{GPT-5.2}~\cite{OpenAI2025},
\textbf{Claude-4.5 Sonnet (no thinking)}~\cite{anthropic_claude_4_5_sonnet2025}, and
\textbf{Qwen3-max}~\cite{qwen3_max_2025}.

Table~\ref{tab:consolidated_abpr} reports the performance gains obtained by embedding these base models within the ABPR framework.
Across all evaluated models, ABPR consistently improves accuracy, indicating
that it acts as a reasoning amplifier rather than a model-specific
optimisation.
The effect is particularly pronounced for models with pre-trained (plausible) reasoning capacity.
For example, \textbf{GPT-5.2} under a low reasoning-effort configuration improves
from a baseline of 8.33\% to \textbf{30.00\%}, while \textbf{Gemini-3-Flash with ABPR} is boosted from 34.03\% to \textbf{56.67\%}.
These results suggest that ABPR effectively augments intrinsic neural reasoning
with external symbolic verification and procedural structure.

Figure~\ref{fig:iter_results} further illustrates this effect by showing
performance as a function of refinement iterations.
Across models, ABPR exhibits steady improvements over successive refinement
steps, indicating that the gains arise from iterative procedural refinement
rather than isolated corrections, which address our RQ1.

Iteration and cost curves are provided in Appendix~\ref{app:iteration_cost_curves}; they show that cumulative solved-task rate increases across refinement steps and report the corresponding cost--performance tradeoff.

\subsection{RQ2: Impact of semantic derivation traces}
\label{subsec:trace_ablation}

This experiment isolates the contribution of proof-tree-style semantic derivations. Specifically, we compare ABPR against execution-feedback repair on Gemini-3 series models. In the baseline setting, the LLM can revise programs iteratively from verifier outcomes and error logs, but receives no APD derivation tree.


The execution-feedback repair rows in Table~\ref{tab:consolidated_abpr} report the
performance of this baseline.
For both \textbf{Gemini-3-Pro} and \textbf{Gemini-3-Flash}, removing declarative traces and the
associated semantic localization strategy results in a substantial drop in
performance, with Pass@2 decreasing by up to 10.00 percentage points.

The results answer RQ2: semantic derivation traces play a central role in guiding the refinement procedure.
Rather than serving as auxiliary context, these traces provide structured intermediate evidence about the goal--subgoal reasoning path. This supports more targeted updates than execution-feedback trial and error, improving refinement efficiency.





\subsection{RQ3: Impact of the quality of initial hypotheses}
\label{subsec:ablation}

Table~\ref{tab:consolidated_abpr} reveals a boundary condition: ABPR is most
effective when the first-pass hypothesis lies within a repairable neighbourhood
of the target rule. With the same Claude ABPR refiner, the independent
Gemini-3-Pro initialization pool used in the cross-model sweep improves from
27.92\% before repair to 42.92\% after Claude refinement, far above Claude-only
ABPR at 5.00\%. Thus procedural refinement is local: stronger initial rules
place the search in a recoverable basin, while poor initial hypotheses may
remain outside the reach of short stochastic refinement trajectories.

\subsection{RQ4: Generalisation to RAVEN-style matrix reasoning}
\label{subsec:raven_generalization}

To test whether ABPR is tied to ARC-specific grid transformations, we further
evaluate it on fill-in-the-blank adaptations of I-RAVEN-X~\citep{camposampiero2025iravenx}
and A-I-RAVEN~\citep{malkinski2025airaven}.  These supplementary non-ARC tasks
require inferring relational rules over symbolic matrix attributes and filling a
hidden entry; candidate answers are hidden from the prompt and used only by the
scoring environment.

ABPR remains effective in this setting.  On I-RAVEN-X, Prolog direct and Prolog
execution-feedback repair solve 34.00\% and 39.60\% of tasks, while Prolog ABPR
solves 88.60\%.  On A-I-RAVEN-style tasks, the corresponding numbers are
28.20\%, 34.00\%, and 93.60\%.  We also include Python direct and Python
execution-feedback repair as stronger no-trace references: they reach
64.00\%/81.60\% on I-RAVEN-X and 74.00\%/88.60\% on A-I-RAVEN-style tasks.
Thus ABPR exceeds the no-trace Python feedback rows under the adapted protocol.
We interpret this as evidence that
semantic derivation traces provide a reusable refinement signal beyond
ARC-specific grid priors and final execution feedback.  Full per-suite solved counts and accuracies are reported in Appendix~\ref{app:raven_settings}.

\section{Related Work and Discussion}
\label{sec:related_work}
LLM self-correction is unreliable without grounded feedback
~\citep{huang2023large,kamoi2024can}.  Tool-augmented and program-aided methods
externalize reasoning to programs or verifiers~\citep{gao2023pal,schick2023toolformer,wei2022chain};
ABPR follows this direction, but studies proof-tree-style semantic derivations
of executable rule hypotheses as the feedback object.

ABPR is related to automated program repair through localization,
verification, and iterative revision.  Recent LLM-based APR systems repair
LLM-generated programs~\citep{fan2023automated}, combine logic-based
localization with LLM repair for ASP submissions~\citep{brancas2025combining},
or use counterexamples and MaxSAT-based localization to constrain patches
~\citep{orvalho2025counterexample}.  These works repair software or
logic-program artifacts; here the program is an executable hypothesis of an
abstract rule.  We therefore compare feedback conditions under matched abstract
reasoning protocols rather than claiming to outperform APR systems on their
benchmarks.  Our supplementary RAVEN settings draw on I-RAVEN-X
~\citep{camposampiero2025iravenx} and A-I-RAVEN~\citep{malkinski2025airaven},
and the overall approach is connected to neuro-symbolic reasoning
~\citep{hitzler2022neuro,DeRaedt26Manifesto,GarcezL23:3rdwave,Manhaeve2018DeepProbLog,dai:abl,BadreddineGSS22:LTN}.

\section{Conclusion}
\label{sec:conclusion}
We presented ABPR, a semantic-trace-guided framework for abstract rule
induction.  ABPR treats candidate programs as executable declarative hypotheses
and uses APD-style proof-tree derivations to convert outcome-level failures into
localized semantic revision problems.  ARC-AGI-2 and supplementary RAVEN-style
experiments show that this refinement signal improves reasoning beyond direct
or execution-feedback repair.

\bibliographystyle{plainnat}
\bibliography{example_paper}

@article{adnan2025measuring,
  title={Measuring and mitigating debugging effectiveness decay in code language models},
  author={Adnan, Muntasir and Kuhn, Carlos CN},
  journal={Scientific Reports},
  volume={15},
  number={1},
  pages={44120},
  year={2025},
  publisher={Nature Publishing Group UK London},
  doi={10.1038/s41598-025-27846-5},
  url={https://doi.org/10.1038/s41598-025-27846-5}
}

@inproceedings{fan2023automated,
  author       = {Zhiyu Fan and
                  Xiang Gao and
                  Martin Mirchev and
                  Abhik Roychoudhury and
                  Shin Hwei Tan},
  title        = {Automated Repair of Programs from Large Language Models},
  booktitle    = {45th {IEEE/ACM} International Conference on Software Engineering,
                  {ICSE} 2023, Melbourne, Australia, May 14-20, 2023},
  pages        = {1469--1481},
  publisher    = {{IEEE}},
  year         = {2023},
  url          = {https://doi.org/10.1109/ICSE48619.2023.00128},
  doi          = {10.1109/ICSE48619.2023.00128},
  bibsource    = {dblp computer science bibliography, https://dblp.org}
}

@inproceedings{brancas2025combining,
  author       = {Ricardo Brancas and
                  Vasco Manquinho and
                  Ruben Martins},
  title        = {Combining Logic and Large Language Models for Assisted Debugging and
                  Repair of {ASP} Programs},
  booktitle    = {{IEEE} Conference on Software Testing, Verification and Validation,
                  {ICST} 2025, Napoli, Italy, March 31 - April 4, 2025},
  pages        = {646--657},
  publisher    = {{IEEE}},
  year         = {2025},
  url          = {https://doi.org/10.1109/ICST62969.2025.10988950},
  doi          = {10.1109/ICST62969.2025.10988950},
  bibsource    = {dblp computer science bibliography, https://dblp.org}
}

@article{orvalho2025counterexample,
  author       = {Pedro Orvalho and
                  Mikol{\'{a}}s Janota and
                  Vasco Manquinho},
  title        = {Counterexample Guided Program Repair Using Zero-Shot Learning and
                  MaxSAT-based Fault Localization},
  journal      = {CoRR},
  volume       = {abs/2502.07786},
  year         = {2025},
  url          = {https://doi.org/10.48550/arXiv.2502.07786},
  doi          = {10.48550/ARXIV.2502.07786},
  eprinttype   = {arXiv},
  eprint       = {2502.07786},
  bibsource    = {dblp computer science bibliography, https://dblp.org}
}

@inproceedings{huang2023large,
  author       = {Jie Huang and
                  Xinyun Chen and
                  Swaroop Mishra and
                  Huaixiu Steven Zheng and
                  Adams Wei Yu and
                  Xinying Song and
                  Denny Zhou},
  title        = {Large Language Models Cannot Self-Correct Reasoning Yet},
  booktitle    = {The Twelfth International Conference on Learning Representations,
                  {ICLR} 2024},
  year         = {2024},
  bibsource    = {dblp computer science bibliography, https://dblp.org}
}

@inproceedings{chen2025studying,
  title={Studying and Understanding the Effectiveness and Failures of Conversational {LLM}-Based Repair},
  author={Aolin Chen and
                  Haojun Wu and
                  Qi Xin and
                  Steven P. Reiss and
                  Jifeng Xuan},
  booktitle={2025 IEEE/ACM International Workshop on Automated Program Repair (APR)},
  pages={56--59},
  year={2025},
  publisher={IEEE},
  doi={10.1109/APR66717.2025.00014},
  url={https://doi.org/10.1109/APR66717.2025.00014}
}

@techreport{rossi2025aaai,
  title={{AAAI} 2025 presidential panel on the future of {AI} research},
  author={Rossi, Francesca and others},
  institution={Association for the Advancement of Artificial Intelligence},
  year={2025},
  url={https://aaai.org/about-aaai/presidential-panel-on-the-future-of-ai-research/},
  note={Presidential panel report}
}

@article{chollet2025arc,
  author       = {Fran{\c{c}}ois Chollet and
                  Mike Knoop and
                  Gregory Kamradt and
                  Bryan Landers and
                  Henry Pinkard},
  title        = {{ARC-AGI}-2: {A} New Challenge for Frontier {AI} Reasoning Systems},
  journal      = {CoRR},
  volume       = {abs/2505.11831},
  year         = {2025},
  doi          = {10.48550/ARXIV.2505.11831},
  eprinttype   = {arXiv},
  eprint       = {2505.11831},
  bibsource    = {dblp computer science bibliography, https://dblp.org}
}

@article{kakas1992abductive,
  title={Abductive logic programming},
  author={Kakas, Antonis C and Kowalski, Robert A. and Toni, Francesca},
  journal={Journal of logic and computation},
  volume={2},
  number={6},
  pages={719--770},
  year={1992},
  publisher={Oxford University Press}
}

@article{muggleton1995inverse,
  title={Inverse entailment and Progol},
  author={Muggleton, Stephen},
  journal={New generation computing},
  volume={13},
  number={3},
  pages={245--286},
  year={1995},
  publisher={Springer}
}

@article{kamoi2024can,
  author       = {Ryo Kamoi and
                  Yusen Zhang and
                  Nan Zhang and
                  Jiawei Han and
                  Rui Zhang},
  title        = {When Can {LLM}s Actually Correct Their Own Mistakes? {A} Critical
                  Survey of Self-Correction of {LLM}s},
  journal      = {Trans. Assoc. Comput. Linguistics},
  volume       = {12},
  pages        = {1417--1440},
  year         = {2024},
  doi          = {10.1162/TACL_A_00713},
  bibsource    = {dblp computer science bibliography, https://dblp.org}
}

@article{zhou2019abductive,
  author       = {Zhi{-}Hua Zhou},
  title        = {Abductive {L}earning: towards bridging machine learning and logical
                  reasoning},
  journal      = {Science China Information Sciences},
  volume       = {62},
  number       = {7},
  pages        = {76101:1--76101:3},
  year         = {2019},
 //url          = {https://doi.org/10.1007/s11432-018-9801-4},
  doi          = {10.1007/S11432-018-9801-4},
  timestamp    = {Mon, 02 Mar 2020 16:31:11 +0100},
  biburl       = {https://dblp.org/rec/journals/chinaf/Zhou19.bib},
  bibsource    = {dblp computer science bibliography, https://dblp.org}
}

@book{shapiro1982algorithmic,
  title={Algorithmic {P}rogram {D}ebugging},
  author={Shapiro, Ehud Yehuda},
  year={1982},
  publisher={Yale University}
}

@misc{Gemini3TechReport2025,
  author       = {{Google}},
  title        = {A New Era of Intelligence with {Gemini} 3},
  howpublished = {Google Keyword Blog},
  year         = {2025},
  month        = {November},
  url          = {https://blog.google/products-and-platforms/products/gemini/gemini-3/}
}

@misc{Poetiq2025Frontier,
  author       = {{Poetiq Team}},
  title        = {Traversing the Frontier of Superintelligence: Poetiq Shatters {ARC-AGI}-2 State of the Art},
  howpublished = {Poetiq AI Technical Blog},
  year         = {2025},
  month        = {January},
  url          = {https://poetiq.ai/posts/arcagi_announcement/}
}

@book{lloyd2012foundations,
  title={Foundations of {L}ogic {P}rogramming},
  author={Lloyd, John W},
  year={2012},
  publisher={Springer Science \& Business Media}
}

@article{robinson1965machine,
  author       = {John Alan Robinson},
  title        = {A Machine-Oriented Logic Based on the Resolution Principle},
  journal      = {J. {ACM}},
  volume       = {12},
  number       = {1},
  pages        = {23--41},
  year         = {1965},
  //url          = {https://doi.org/10.1145/321250.321253},
  doi          = {10.1145/321250.321253},
  timestamp    = {Fri, 24 Mar 2023 16:31:07 +0100},
  biburl       = {https://dblp.org/rec/journals/jacm/Robinson65.bib},
  bibsource    = {dblp computer science bibliography, https://dblp.org}
}

@book{o2009craft,
  title={The craft of Prolog},
  author={O'Keefe, Richard},
  year={2009},
  publisher={MIT press}
}

@techreport{OpenAI2025,
  author      = {{OpenAI}},
  title       = {{GPT}-5 System Card},
  institution = {OpenAI},
  year        = {2025},
  month       = {August},
  url         = {https://openai.com/index/gpt-5-system-card/}
}

@inproceedings{Manhaeve2018DeepProbLog,
  author       = {Robin Manhaeve and
                  Sebastijan Dumancic and
                  Angelika Kimmig and
                  Thomas Demeester and
                  Luc De Raedt},
  title        = {DeepProbLog: Neural Probabilistic Logic Programming},
  booktitle    = {Advances in Neural Information Processing Systems 31:
                  Annual Conference on Neural Information Processing Systems 2018,
                  NeurIPS 2018},
  pages        = {3753--3763},
  year         = {2018},
  bibsource    = {dblp computer science bibliography, https://dblp.org}
}

@inproceedings{holtzman2019curious,
  author       = {Ari Holtzman and
                  Jan Buys and
                  Li Du and
                  Maxwell Forbes and
                  Yejin Choi},
  title        = {The Curious Case of Neural Text Degeneration},
  booktitle    = {8th International Conference on Learning Representations,
                  {ICLR} 2020},
  year         = {2020},
  bibsource    = {dblp computer science bibliography, https://dblp.org}
}

@article{chen2021evaluating,
  author       = {Mark Chen and
                  Jerry Tworek and
                  Heewoo Jun and
                  Qiming Yuan and
                  Henrique Ponde de Oliveira Pinto and
                  Jared Kaplan and
                  Harri Edwards and
                  Yuri Burda and
                  Nicholas Joseph and
                  Greg Brockman and
                  Alex Ray and
                  Ra{\'{u}}l Puri and
                  Gretchen Krueger and
                  Michael Petrov and
                  Heidy Khlaaf and
                  Girish Sastry and
                  Pamela Mishkin and
                  Brooke Chan and
                  Scott Gray and
                  Nick Ryder and
                  Mikhail Pavlov and
                  Alethea Power and
                  Lukasz Kaiser and
                  Mohammad Bavarian and
                  Clemens Winter and
                  Philippe Tillet and
                  Felipe Petroski Such and
                  Dave Cummings and
                  Matthias Plappert and
                  Fotios Chantzis and
                  Elizabeth Barnes and
                  Ariel Herbert{-}Voss and
                  William Hebgen Guss and
                  Alex Nichol and
                  Alex Paino and
                  Nikolas Tezak and
                  Jie Tang and
                  Igor Babuschkin and
                  Suchir Balaji and
                  Shantanu Jain and
                  William Saunders and
                  Christopher Hesse and
                  Andrew N. Carr and
                  Jan Leike and
                  Josh Achiam and
                  Vedant Misra and
                  Evan Morikawa and
                  Alec Radford and
                  Matthew Knight and
                  Miles Brundage and
                  Mira Murati and
                  Katie Mayer and
                  Peter Welinder and
                  Bob McGrew and
                  Dario Amodei and
                  Sam McCandlish and
                  Ilya Sutskever and
                  Wojciech Zaremba},
  title        = {Evaluating Large Language Models Trained on Code},
  journal      = {CoRR},
  volume       = {abs/2107.03374},
  year         = {2021},
  eprinttype   = {arXiv},
  eprint       = {2107.03374},
  bibsource    = {dblp computer science bibliography, https://dblp.org}
}

@inproceedings{dietterich2000ensemble,
  title={Ensemble methods in {M}achine {L}earning},
  author={Dietterich, Thomas G},
  booktitle={Multiple Classifier Systems, First International Workshop},
  pages={1--15},
  year={2000},
  address = { Cagliari, Italy},
  organization={Springer}
}

@article{kuncheva2003measures,
  title={Measures of diversity in classifier ensembles and their relationship with the ensemble accuracy},
  author={Kuncheva, Ludmila I and Whitaker, Christopher J},
  journal={Machine learning},
  volume={51},
  number={2},
  pages={181--207},
  year={2003},
  publisher={Springer}
}

@misc{anthropic_claude_4_5_sonnet2025,
  title        = {Introducing {Claude} Sonnet 4.5},
  author       = {{Anthropic}},
  year         = {2025},
  month        = {September},
  howpublished = {Anthropic News},
  url          = {https://www.anthropic.com/news/claude-sonnet-4-5}
}

@misc{qwen3_max_2025,
  title        = {{Qwen3-Max}: Just Scale it},
  author       = {{Qwen Team}},
  year         = {2025},
  month        = {September},
  howpublished = {Qwen Blog},
  url          = {https://qwen.ai/blog?from=home.latest-research-list&id=87dc93fc8a590dc718c77e1f6e84c07b474f6c5a}
}

@book{hitzler2022neuro,
  editor={Hitzler, Pascal and Sarker, Md Kamruzzaman},
  title={Neuro-Symbolic Artificial Intelligence: The State of the Art},
  series={Frontiers in Artificial Intelligence and Applications},
  volume={342},
  year={2021},
  publisher={IOS Press},
  isbn={978-1-64368-244-0},
  doi={10.3233/FAIA342},
  url={https://doi.org/10.3233/FAIA342}
}

@article{brown2020language,
  title={Language models are few-shot learners},
  author={Brown, Tom and Mann, Benjamin and Ryder, Nick and Subbiah, Melanie and Kaplan, Jared D and Dhariwal, Prafulla and Neelakantan, Arvind and Shyam, Pranav and Sastry, Girish and Askell, Amanda and others},
  journal={Advances in neural information processing systems},
  volume={33},
  pages={1877--1901},
  year={2020}
}

@article{wei2022chain,
  author       = {Jason Wei and
                  Xuezhi Wang and
                  Dale Schuurmans and
                  Maarten Bosma and
                  Ed H. Chi and
                  Quoc Le and
                  Denny Zhou},
  title        = {Chain of Thought Prompting Elicits Reasoning in Large Language Models},
  journal      = {CoRR},
  volume       = {abs/2201.11903},
  year         = {2022},
  //url          = {https://arxiv.org/abs/2201.11903},
  eprinttype    = {arXiv},
  eprint       = {2201.11903},
  timestamp    = {Fri, 22 Apr 2022 16:06:31 +0200},
  biburl       = {https://dblp.org/rec/journals/corr/abs-2201-11903.bib},
  bibsource    = {dblp computer science bibliography, https://dblp.org}
}

@inproceedings{schick2023toolformer,
   author       = {Timo Schick and
                  Jane Dwivedi{-}Yu and
                  Roberto Dess{\`{\i}} and
                  Roberta Raileanu and
                  Maria Lomeli and
                  Eric Hambro and
                  Luke Zettlemoyer and
                  Nicola Cancedda and
                  Thomas Scialom},
  title        = {Toolformer: Language Models Can Teach Themselves to Use Tools},
  booktitle    = {Advances in Neural Information Processing Systems 36},
  year         = {2023},
  address = {New Orleans, LA},
  //url          = {http://papers.nips.cc/paper\_files/paper/2023/hash/d842425e4bf79ba039352da0f658a906-Abstract-Conference.html},
  timestamp    = {Fri, 01 Mar 2024 16:26:21 +0100},
  biburl       = {https://dblp.org/rec/conf/nips/SchickDDRLHZCS23.bib},
  bibsource    = {dblp computer science bibliography, https://dblp.org}
}

@inproceedings{gao2023pal,
  author       = {Luyu Gao and
                  Aman Madaan and
                  Shuyan Zhou and
                  Uri Alon and
                  Pengfei Liu and
                  Yiming Yang and
                  Jamie Callan and
                  Graham Neubig},
  title        = {{PAL:} Program-Aided Language Models},
  booktitle    = {International Conference on Machine Learning, {ICML} 2023},
  series       = {Proceedings of Machine Learning Research},
  volume       = {202},
  pages        = {10764--10799},
  publisher    = {{PMLR}},
  year         = {2023},
  bibsource    = {dblp computer science bibliography, https://dblp.org}
}

@article{cobbe2021training,
  author       = {Karl Cobbe and
                  Vineet Kosaraju and
                  Mohammad Bavarian and
                  Mark Chen and
                  Heewoo Jun and
                  Lukasz Kaiser and
                  Matthias Plappert and
                  Jerry Tworek and
                  Jacob Hilton and
                  Reiichiro Nakano and
                  Christopher Hesse and
                  John Schulman},
  title        = {Training Verifiers to Solve Math Word Problems},
  journal      = {CoRR},
  volume       = {abs/2110.14168},
  year         = {2021},
  eprinttype   = {arXiv},
  eprint       = {2110.14168},
  bibsource    = {dblp computer science bibliography, https://dblp.org}
}

@article{saunders2022self,
  author       = {William Saunders and
                  Catherine Yeh and
                  Jeff Wu and
                  Steven Bills and
                  Long Ouyang and
                  Jonathan Ward and
                  Jan Leike},
  title        = {Self-Critiquing Models for Assisting Human Evaluators},
  journal      = {CoRR},
  volume       = {abs/2206.05802},
  year         = {2022},
  eprinttype   = {arXiv},
  eprint       = {2206.05802},
  bibsource    = {dblp computer science bibliography, https://dblp.org}
}

@inproceedings{lightman2023lets,
  author       = {Hunter Lightman and
                  Vineet Kosaraju and
                  Yuri Burda and
                  Harrison Edwards and
                  Bowen Baker and
                  Teddy Lee and
                  Jan Leike and
                  John Schulman and
                  Ilya Sutskever and
                  Karl Cobbe},
  title        = {Let's Verify Step by Step},
  booktitle    = {The 12th International Conference on Learning Representations},
    address = {Vienna, Austria},
  year         = {2024},
  //crossref     = {DBLP:conf/iclr/2024},
  //url          = {https://openreview.net/forum?id=v8L0pN6EOi},
  timestamp    = {Wed, 07 Aug 2024 17:11:53 +0200},
  biburl       = {https://dblp.org/rec/conf/iclr/LightmanKBEBLLS24.bib},
  bibsource    = {dblp computer science bibliography, https://dblp.org}
}

@inproceedings{gou2024critic,
  author       = {Zhibin Gou and
                  Zhihong Shao and
                  Yeyun Gong and
                  Yelong Shen and
                  Yujiu Yang and
                  Nan Duan and
                  Weizhu Chen},
  title        = {{CRITIC:} Large Language Models Can Self-Correct with
                  Tool-Interactive Critiquing},
  booktitle    = {The Twelfth International Conference on Learning Representations,
                  {ICLR} 2024},
  year         = {2024},
  bibsource    = {dblp computer science bibliography, https://dblp.org}
}

@article{ridnik2024alphacodium,
   author       = {Tal Ridnik and
                  Dedy Kredo and
                  Itamar Friedman},
  title        = {Code Generation with AlphaCodium: From Prompt Engineering to Flow
                  Engineering},
  journal      = {CoRR},
  volume       = {abs/2401.08500},
  year         = {2024},
  //url          = {https://doi.org/10.48550/arXiv.2401.08500},
  doi          = {10.48550/ARXIV.2401.08500},
  eprinttype    = {arXiv},
  eprint       = {2401.08500},
  timestamp    = {Thu, 01 Feb 2024 15:35:36 +0100},
  biburl       = {https://dblp.org/rec/journals/corr/abs-2401-08500.bib},
  bibsource    = {dblp computer science bibliography, https://dblp.org}
}

@Inbook{DeRaedt26Manifesto,
author={De Raedt, Luc
and Heintz, Fredrik
and Kersting, Kristian
and Marra, Giuseppe},
title={A Learning and Reasoning Manifesto},
bookTitle={Challenges and Algorithms for Knowledge Discovery from Data: Essays Dedicated to Arno Siebes on the Occasion of His 67th Birthday},
year={2025},
publisher={Springer Nature},
address={Cham, Switzerland},
pages={97--108},
doi={10.1007/978-3-032-03028-3_6},
url={https://doi.org/10.1007/978-3-032-03028-3_6}
}

@article{BadreddineGSS22:LTN,
  author       = {Badreddine, Samy and
                  {d'A}vila Garcez, Artur S. and
                  Serafini, Luciano and
                  Spranger, Michael},
  title        = {Logic Tensor Networks},
  journal      = {Artificial Intelligence},
  volume       = {303},
  pages        = {103649},
  year         = {2022},
}

@article{GarcezL23:3rdwave,
  author       = {Garcez, Artur S. d'Avila and Lamb, Luis C.},
  title        = {Neurosymbolic {AI:} the 3rd wave},
  journal      = {Artificial Intelligence Review},
  volume       = {56},
  number       = {11},
  pages        = {12387--12406},
  year         = {2023}
}

@incollection{dai:abl,
  author={Dai, Wang{-}Zhou and Xu, Qiu{-}Ling and Yu, Yang and Zhou, Zhi{-}Hua},
  title={Bridging Machine Learning and Logical Reasoning by Abductive Learning},
  year={2019},
  pages={2811-2822},
  booktitle = {Advances in Neural Information Processing Systems 32},
  publisher = {Curran Associates, Inc.},
}

@article{WielemakerSTL12:SWI,
  author       = {Wielemaker, Jan and
                  Schrijvers, Tom and
                  Triska, Markus and
                  Lager, Torbj{\"{o}}rn},
  title        = {{SWI}-Prolog},
  journal      = {Theory and Practice of Logic Programming},
  volume       = {12},
  number       = {1-2},
  pages        = {67--96},
  year         = {2012}
}

@misc{camposampiero2025iravenx,
  author       = {Camposampiero, Giacomo and Hersche, Michael and Wattenhofer, Roger and Sebastian, Abu and Rahimi, Abbas},
  title        = {{I-RAVEN-X}: Benchmarking Generalization and Robustness of Analogical and Mathematical Reasoning in Large Language and Reasoning Models},
  year         = {2025},
  eprint       = {2510.17496},
  archivePrefix= {arXiv},
  primaryClass = {cs.AI}
}

@inproceedings{malkinski2025airaven,
  author    = {Malkinski, Mikolaj and Mandziuk, Jacek},
  title     = {{A-I-RAVEN} and {I-RAVEN-Mesh}: Two New Benchmarks for Abstract Visual Reasoning},
  booktitle = {Proceedings of the Thirty-Fourth International Joint Conference on Artificial Intelligence},
  year      = {2025},
  pages     = {5932--5940},
  doi       = {10.24963/ijcai.2025/660}
}

\newpage
\appendix
\clearpage

\section{ARC Iteration and Cost Curves}
\label{app:iteration_cost_curves}
\begin{figure}[h]
\centering
\begin{minipage}[t]{0.47\linewidth}
    \centering
    \includegraphics[width=0.94\linewidth]{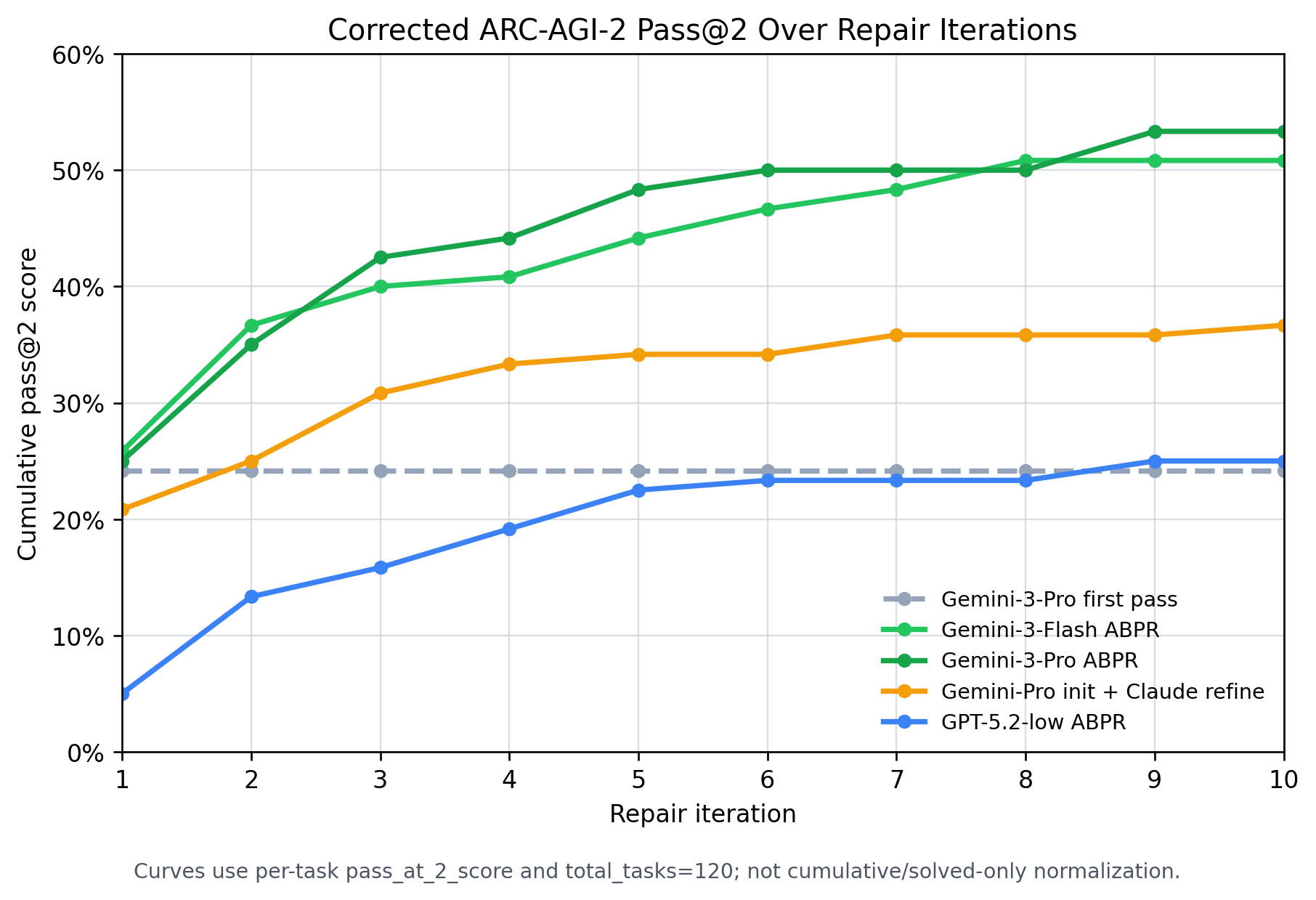}
    \captionof{figure}{Representative cumulative solved-task rate over ABPR refinement iterations.}
    \label{fig:iter_results}
\end{minipage}\hspace{0.06\linewidth}%
\begin{minipage}[t]{0.47\linewidth}
	\centering
	\includegraphics[width=0.94\linewidth]{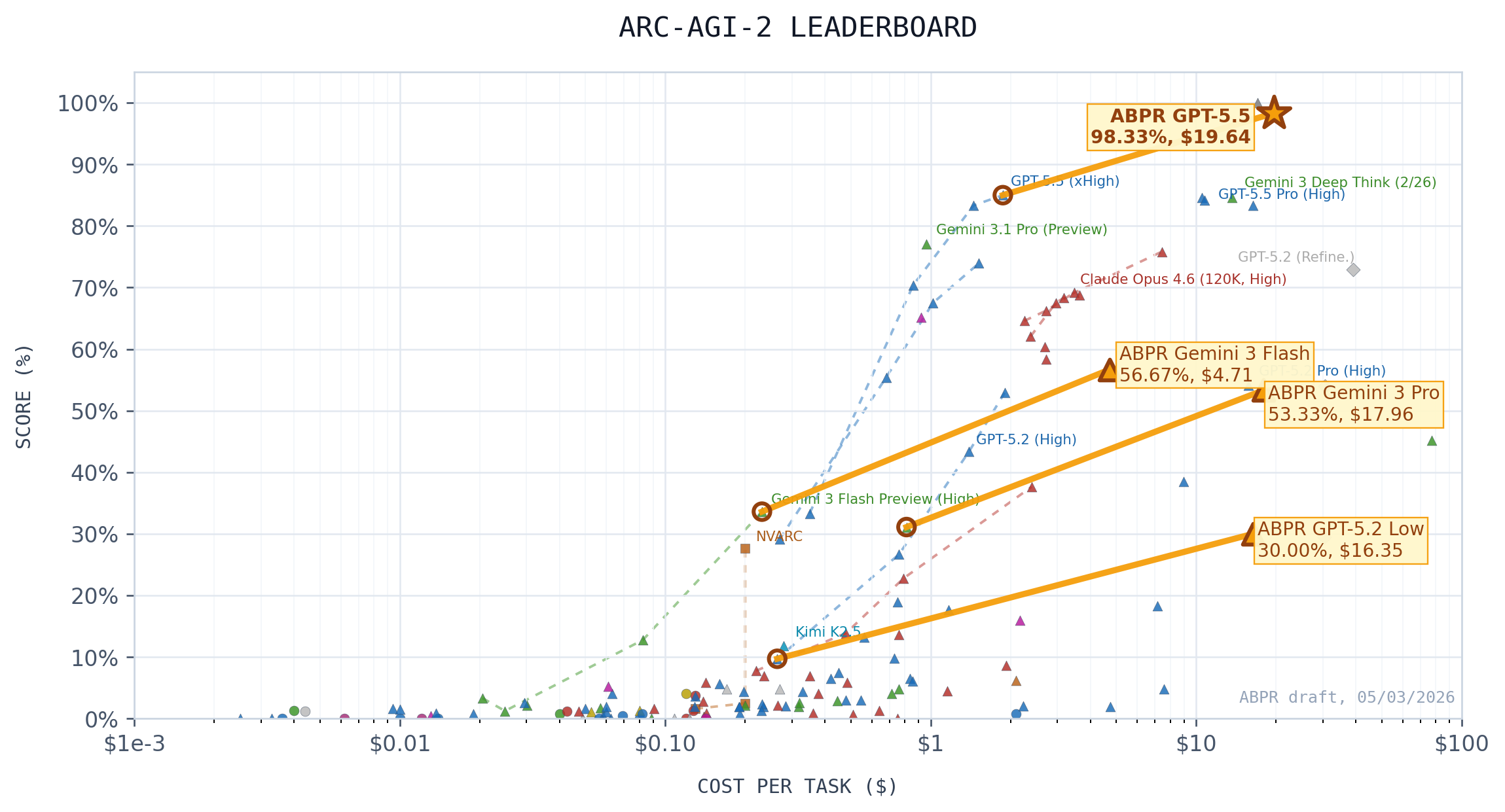}
     \captionof{figure}{Cost and performance of current models on ARC-AGI-2, with ABPR points connected to their leaderboard baselines.}
     \label{fig:cost-score}
\end{minipage}
\end{figure}
\section{Experiment Settings}
\label{app:experiment_settings}

In this section, we provide detailed configurations for reproducibility.

\subsection{Model Configurations}
We evaluate our method on several state-of-the-art Large Language Models (LLMs). All models were accessed via their official APIs.
\begin{itemize}
    \item \textbf{ARC sampling temperature:} For ARC-AGI-2 experiments, we set the
sampling temperature to $T_{\mathrm{sample}}=1.0$ for all models to encourage
exploration during the sampling and refinement phases. RAVEN-style experiments
use the separate setting reported in Appendix~\ref{app:raven_settings}.
    \item \textbf{Reasoning Effort:} For the GPT-series models reported in the main ARC table, we use the low reasoning-effort configuration by explicitly setting \texttt{reasoning\_effort='low'} via the API.
    \item \textbf{Other Parameters:} For all other models (including Gemini-3 series, Claude series, and Qwen series), we utilized the standard ChatCompletion mode with default reasoning configurations provided by the respective official APIs.
\end{itemize}

\subsection{Parallel Search and Evaluation Protocol}
\begin{itemize}
    \item \textbf{Ensemble Strategy ($N=8$):} To robustly solve the tasks, we deploy $N=8$ parallel ABPR processes. Each process operates independently to explore the hypothesis space.
    \item \textbf{Selection Mechanism:} Instead of random sampling, we employ a diversity-prioritised voting mechanism over the $N=8$ final hypotheses to select the top-$k$ ($k=2$) most promising programs.
    \item \textbf{Pass@2 Evaluation:} The reported Pass@2 metric strictly follows the ARC-AGI evaluation protocol: a task is considered solved if at least one of the top-2 selected programs (from the voting stage) produces the correct output. This reflects the practical system performance under the two-submission constraint, rather than a statistical estimate from random sampling.
    \item \textbf{Search Depth:} We use a total-depth convention:
$T_{\mathrm{depth}}=1$ denotes the initial generation only, and
$T_{\mathrm{depth}}>1$ includes $T_{\mathrm{depth}}-1$ subsequent repair
rounds. The main ARC-AGI-2 setting uses $T_{\mathrm{depth}}=10$, i.e., one
initial generation plus up to nine repair rounds. For ablation studies marked
as ``Iteration=1'', the refinement loop is disabled.
\end{itemize}

\subsection{Prompting and Randomness}
\begin{itemize}
    \item \textbf{Prompt Construction:} We utilize few-shot prompting where demonstration examples are retrieved from the training split.
    \item \textbf{Random Seeding:} Each of the $N=8$ parallel processes is initialised with a distinct random seed. This seed controls the permutation of few-shot examples in the context window to induce search diversity across the ensemble, preventing the parallel threads from collapsing into identical refinement trajectories.
\end{itemize}

\subsection{ARC Reproducibility and Variance Analysis}
\label{app:variance_reproducibility}

The ARC-AGI-2 experiments deliberately use stochastic sampling. We set
the sampling temperature to $T_{\mathrm{sample}}=1.0$ and use different random
seeds across parallel threads so that the first-pass generator explores a
broader distribution of candidate programs. The evaluation is therefore not
intended to make each single trajectory deterministic; instead, reproducibility
is obtained by aggregating sufficiently many independent samples across ensemble
size $N$ and total search depth $T_{\mathrm{depth}}$.

\paragraph{Variance metrics.}
For a fixed setting, let $s_r$ denote the Pass@2 score, in percentage points, of
repeat $r$. We report the repeated-run mean
$\bar{s}=\frac{1}{R}\sum_{r=1}^{R}s_r$ and sample standard deviation
$\sqrt{\frac{1}{R-1}\sum_{r=1}^{R}(s_r-\bar{s})^2}$. For the heatmap analysis,
we also report the repeated-run range $\max_r s_r-\min_r s_r$. Lower standard
deviation and lower range indicate higher reproducibility.

\begin{figure}[t]
    \centering
    \includegraphics[width=\linewidth]{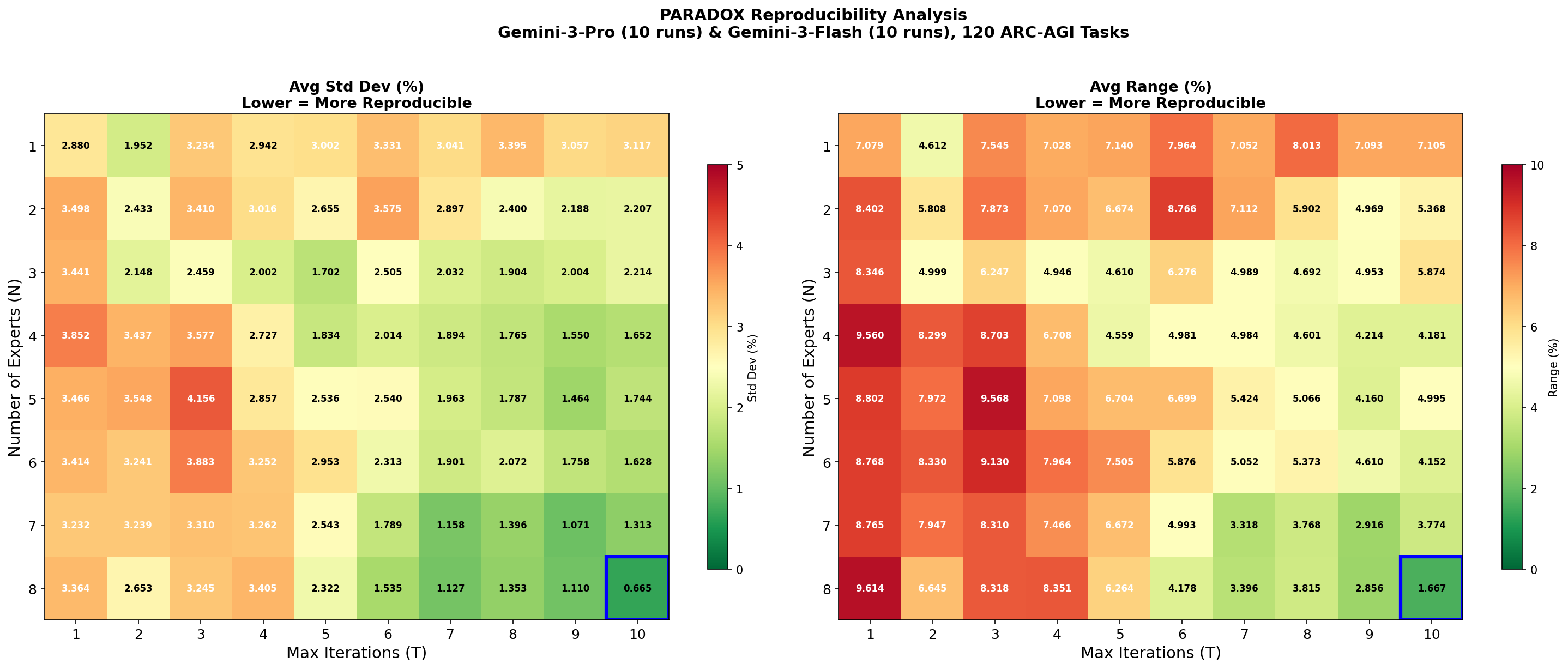}
    \caption{ARC-AGI-2 reproducibility heatmap. Each cell is averaged over
Gemini-3-Pro and Gemini-3-Flash after 10 repeated runs per model at the
corresponding ensemble size $N$ and total search depth $T_{\mathrm{depth}}$.
Here $T_{\mathrm{depth}}=1$ denotes first-pass generation only. Lower values
indicate more stable repeated performance.}
    \label{fig:reproducibility_heatmap}
\end{figure}

\paragraph{Evaluation data and repeated-run protocol.}
The reproducibility heatmaps are computed on the same 120-task ARC-AGI-2 public
evaluation set used in the main experiments. For Gemini-3-Pro and Gemini-3-Flash, we repeat the experiment 10 times for each
pair of ensemble size $N$ and total search depth $T_{\mathrm{depth}}$, then
compute the standard deviation and range for each model and average the two
values cell-wise. All reported non-Gemini ARC result means and standard
deviations are computed from three independent repeated runs under the
corresponding setting. Table~\ref{tab:arc_variance_summary} summarises the
aggregate repeated-run statistics, while Figure~\ref{fig:reproducibility_heatmap}
reports the separate 10-run Gemini reproducibility sweeps over
$(N,T_{\mathrm{depth}})$.

\begin{table}[t]
\centering
\caption{Repeated-run ARC-AGI-2 Pass@2 variance summary. Scores are percentages; Std. is the sample standard deviation across repeated runs.}
\label{tab:arc_variance_summary}
\scriptsize
\setlength{\tabcolsep}{4pt}
\begin{tabular*}{0.72\linewidth}{@{\extracolsep{\fill}}lcc@{}}
\toprule
\textbf{Setting} & \textbf{Mean} & \textbf{Std.} \\
\midrule
Gemini-3-Flash ABPR & 56.25 & 0.59 \\
Gemini-3-Pro ABPR & 52.29 & 1.05 \\
GPT-5.2 low-reasoning ABPR & 29.93 & 0.98 \\
Claude-Sonnet-4.5 ABPR & 4.08 & 0.37 \\
Qwen3-Max ABPR & 2.77 & 0.78 \\
GPT-5.5 xHigh ABPR & 98.05 & 0.48 \\
\bottomrule
\end{tabular*}
\end{table}

Figure~\ref{fig:reproducibility_heatmap} reports the full reproducibility
heatmap over $N$ and $T_{\mathrm{depth}}$. The left panel is the cell-wise
averaged standard deviation and the right panel is the cell-wise averaged range,
computed from the 10 repeated Gemini-3-Pro and Gemini-3-Flash sweeps. The
lower-right region shows that larger $N$ and $T_{\mathrm{depth}}$ substantially
reduce run-to-run variability; at $N=8,T_{\mathrm{depth}}=10$, the averaged
standard deviation is 0.665 percentage points and the averaged range is 1.667
percentage points. This supports the design choice of
using sufficient parallel sampling and iterative refinement rather than relying
on a single stochastic trajectory.

\subsection{RAVEN-style Generalisation Experiments}
\label{app:raven_settings}

The RAVEN-style experiments in Table~\ref{tab:raven_generalization_full} are adapted
to the same verifier-in-the-loop refinement protocol as the ARC-AGI-2 experiments. They are
not evaluated under the original visual candidate-selection setup. Instead, the
model must write or repair a solver that infers the hidden matrix entry
from visible symbolic facts.

\begin{itemize}
    \item \textbf{I-RAVEN-X interactive suite:} We evaluate on 500 official
    I-RAVEN-X test-split symbolic matrix instances adapted to an interactive
    fill-in format. Each instance uses $n=10$, $n_{\text{show}}=3$,
    \texttt{maxval}=1000, and no confounding attributes in the reported final
    suite. Candidate answers are available to the runner only for scoring and
    are hidden from the model prompt.
    \item \textbf{AIRAVEN-style suite:} We evaluate on a 500-query symbolic
    suite derived from I-RAVEN-X primitives and A-I-RAVEN-style
    held-out-attribute regimes. The suite is grouped across ten held-out
    regimes, with 4 visible training puzzles and 10 hidden puzzles per
    generated task group. The generator uses $n=10$, $n_{\text{show}}=3$,
    \texttt{maxval}=1000, \texttt{nconf}=0, and seed 20260501.
    \item \textbf{Backbone and sampling:} All reported RAVEN-style Prolog
    experiments use the same Qwen3-Coder backend, temperature $T=0.5$, and
    20-way worker parallelism.
    \item \textbf{Compared methods:} \emph{Prolog direct} performs a single
    program-generation attempt. \emph{Execution-feedback repair} starts from the
    direct result and refines using final execution feedback only.
    \emph{ABPR repair} starts from the same direct result but additionally uses
    declarative trace-based oracle guidance and local procedural repair.
    \item \textbf{ABPR limits:} For the reported RAVEN-style runs, ABPR is
    limited to at most two refinement rounds and at most five oracle questions
    per task instance. These limits are intentionally small to test whether
    semantic traces improve the local refinement process rather than relying
    on unlimited resampling.
    \item \textbf{Evaluation metric:} We report exact hidden-query accuracy.
    A prediction is correct only if the produced Prolog solver returns the
    withheld expected answer for the hidden matrix entry. Candidate identities
    are used by the runner to score the answer, but are not exposed as a
    candidate-ranking prompt.
\end{itemize}

\begin{table}[h]
\centering
\caption{Full RAVEN-style symbolic matrix results.}
\label{tab:raven_generalization_full}
\scriptsize
\setlength{\tabcolsep}{3.0pt}
\begin{tabular*}{\linewidth}{@{\extracolsep{\fill}}lrrrr@{}}
\toprule
\multirow{2}{*}{\textbf{Method}} &
\multicolumn{2}{c}{\textbf{I-RAVEN-X interactive}} &
\multicolumn{2}{c}{\textbf{A-I-RAVEN-style}} \\
\cmidrule(lr){2-3}\cmidrule(lr){4-5}
& \textbf{Solved} & \textbf{Acc.}
& \textbf{Solved} & \textbf{Acc.} \\
\midrule
Prolog direct & 170/500 & 34.00\% & 141/500 & 28.20\% \\
Prolog execution-feedback repair & 198/500 & 39.60\% & 170/500 & 34.00\% \\
Prolog ABPR repair & 443/500 & \textbf{88.60\%} & 468/500 & \textbf{93.60\%} \\
\midrule
Python direct & 320/500 & 64.00\% & 370/500 & 74.00\% \\
Python execution-feedback repair & 408/500 & 81.60\% & 443/500 & 88.60\% \\
\bottomrule
\end{tabular*}
\end{table}

\section{Example: Execution Debugging vs. Semantic Re-checking}
\label{app:semantic_example}

Consider a graph-reachability rule over the chain
\(a \rightarrow b \rightarrow c \rightarrow d\).  The intended abstraction is
transitive closure: \texttt{reachable(a,d)} should hold.  A Python
implementation may instead encode only two-hop reachability:
\begin{verbatim}
def reachable(x, y):
    if edge(x, y):
        return True
    for z in successors(x):
        if edge(z, y):      # should recurse on reachable(z, y)
            return True
    return False
\end{verbatim}
This program executes normally.  A breakpoint debugger can show the concrete
state \texttt{z=b} and the failed check \texttt{edge(b,d)}, but the semantic
problem is that \texttt{reachable} has been defined as direct-or-two-hop
reachability rather than transitive closure.

In Prolog, the same erroneous hypothesis can be written as:
\begin{verbatim}
reachable(X, Y) :- edge(X, Y).
reachable(X, Y) :- edge(X, Z), edge(Z, Y).
\end{verbatim}
For the query \texttt{reachable(a,d)}, a meta-interpreter reifies the failed
goal--subgoal derivation:
\begin{verbatim}
reachable(a,d)
  -> edge(a,d) fails
  -> edge(a,b) succeeds
  -> edge(b,d) fails
\end{verbatim}
The derivation exposes the rule-level error: the second clause tests
\texttt{edge(Z,Y)} instead of the recursive subgoal
\texttt{reachable(Z,Y)}.  ABPR therefore revises the semantic definition of the
hypothesised relation, rather than patching a single observed execution.

\section{Declarative Trace Meta-Interpreter}
\label{app:trace_meta_interpreter}

ABPR reifies proof-tree-style semantic derivations with a Prolog meta-interpreter rather than by
asking the model to narrate its own execution. The implementation used in our
experiments builds proof-tree nodes of the form
\texttt{node(Type, InputSnapshot, OutputSnapshot, Children)}. A node therefore
records the predicate being resolved, the bindings before and after resolution,
and the recursively selected sub-goals that explain the produced output. The
full implementation additionally imposes a maximum trace depth, suppresses
uninformative system and background-knowledge predicates, expands selected
domain-specific calls such as \texttt{bk:map\_grid\_cells/3}, and prints the
result as a compact declarative proof tree. The core functionality is shown
below.

\begin{lstlisting}[style=appendixprolog]
% Core excerpt of the SWI-Prolog trace meta-interpreter.
% Full code also includes depth limits, output summarisation,
% background-predicate filtering, and formatted proof-tree printing.

solve_and_trace(Goal, TraceTree) :-
    solve_with_depth(Goal, 0, TraceTree).

solve_with_depth(Goal, Depth, TraceTree) :-
    max_trace_depth(MaxDepth),
    ( Depth > MaxDepth ->
        copy_term(Goal, Input),
        call(Goal),
        copy_term(Goal, Output),
        TraceTree = node(depth_limit, Input, Output, [])
    ; interpret_goal(Goal, Depth, TraceTree)
    ).

interpret_goal(true, _Depth, node(builtin, true, true, [])) :- !.
interpret_goal(fail, _Depth, _) :- !, fail.
interpret_goal(!, _Depth, node(builtin, '!', '!', [])) :- !.

interpret_goal((A, B), Depth,
               node(conj, (A, B), (A1, B1), [TreeA|TreesB])) :- !,
    solve_with_depth(A, Depth, TreeA),
    TreeA = node(_, _, A1, _),
    interpret_conjunction_rest(B, Depth, TreesB, B1).

interpret_goal((A ; B), Depth, TraceTree) :- !,
    ( solve_with_depth(A, Depth, TraceTree)
    ; solve_with_depth(B, Depth, TraceTree)
    ).

interpret_goal((Cond -> Then ; Else), Depth, TraceTree) :- !,
    ( solve_with_depth(Cond, Depth, CondTree) ->
        solve_with_depth(Then, Depth, ThenTree),
        ThenTree = node(_, _, Result, _),
        TraceTree = node(if_then_else,
                         (Cond -> Then ; Else),
                         Result,
                         [CondTree, ThenTree])
    ; solve_with_depth(Else, Depth, TraceTree)
    ).

interpret_goal(\+(A), _Depth, node(negation, \+(A), success, [])) :- !,
    \+ call(A).

interpret_goal(findall(Template, Goal, List), _Depth,
               node(findall,
                    findall(TemplateCopy, GoalSummary, _),
                    findall(TemplateCopy, GoalSummary, ListSummary),
                    [])) :- !,
    copy_term(Template, TemplateCopy),
    summarize_goal(Goal, GoalSummary),
    findall(Template, call(Goal), List),
    summarize_list_result(List, ListSummary).

interpret_goal(Goal, Depth,
               node(Functor/Arity, InputSnap, OutputSnap, Children)) :-
    functor(Goal, Functor, Arity),
    copy_term(Goal, InputSnap),
    can_access_clauses(Goal),
    clause(Goal, Body),
    Depth1 is Depth + 1,
    expand_body(Body, Depth1, Children),
    copy_term(Goal, OutputSnap).

expand_body(true, _Depth, []) :- !.
expand_body((A, B), Depth, Children) :- !,
    expand_body(A, Depth, ChildrenA),
    expand_body(B, Depth, ChildrenB),
    append(ChildrenA, ChildrenB, Children).
expand_body(Goal, Depth, [Tree]) :-
    should_trace_goal(Goal),
    solve_with_depth(Goal, Depth, Tree).
expand_body(Goal, _Depth, []) :-
    call(Goal).
\end{lstlisting}

\section{Working examples}

\subsection{Task b0039139}

\begin{figure}[b]
    \centering
    \includegraphics[width=0.92\linewidth]{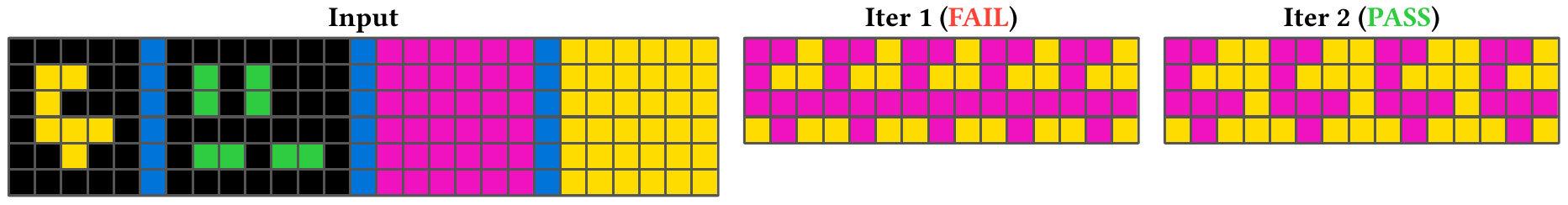} 
    \caption{An iteration example for task b0039139.}
    \label{fig:arc_example_appendix1}
    \vspace{-1em}
\end{figure}
Code of Iteration 1:
\begin{tcolorbox}[  
listing only,  
width=\linewidth,  
breakable,  
colframe=black,  
colback=white,  
boxrule=0.5pt,  
sharp corners,  
left=2mm,  
right=2mm,  
top=.5mm,  
bottom=.5mm
]
\begin{lstlisting}
:- use_module(bk).

solve(Input, Output) :-
    % 1. Determine if separators are horizontal or vertical
    (   is_horizontal(Input)
    ->  bk:grid_dimensions(Input, H, W),
        findall(R, (between(1, H, R), is_separator_row(Input, R)), Seps),
        split_horizontal(Input, Seps, Regions),
        maplist(transpose_grid, Regions, TRegions),
        Transposed = true
    ;   bk:grid_dimensions(Input, H, W),
        findall(C, (between(1, W, C), is_separator_col(Input, C)), Seps),
        split_vertical(Input, Seps, TRegions),
        Transposed = false
    ),
    
    % 2. Extract R1, R2, R3, R4
    TRegions = [R1, R2, R3, R4],
    
    % 3. Extract Kernel K from R1 (color 4)
    bk:connected_components(R1, 4, KComps),
    extract_all_cells(KComps, KCells),
    bk:component_bbox(KCells, bbox(MinR1, MaxR1, MinC1, MaxC1)),
    HK is MaxR1 - MinR1 + 1,
    WK is MaxC1 - MinC1 + 1,
    
    % 4. Extract Points S from R2 (color 3)
    bk:connected_components(R2, 3, SComps),
    extract_all_cells(SComps, SCells),
    findall(C, member((_, C), SCells), SCols),
    sort(SCols, SortedSCols),
    
    % 5. Extract target colors C3 and C4
    extract_block_color(R3, C3),
    extract_block_color(R4, C4),
    
    % 6. Calculate output dimensions and kernel positions
    (   SortedSCols = [] -> Output = [] ;
        min_list(SortedSCols, MinSC),
        calculate_positions(SortedSCols, MinSC, WK, 0, IndexedPositions, MaxPos),
        OutW is MaxPos + WK - 1,
        OutH is HK,
        
        % 7. Create and fill output grid
        length(EmptyRows, OutH),
        maplist(fill_row_with_dummy(OutW), EmptyRows),
        bk:map_grid_cells(EmptyRows, mapper(IndexedPositions, R1, MinR1, MinC1, WK, C3, C4), FilledGrid),
        
        % 8. Final Transpose if necessary
        (   Transposed = true
        ->  transpose_grid(FilledGrid, Output)
        ;   Output = FilledGrid
        )
    ).

% Helper predicates

is_horizontal(Grid) :-
    bk:grid_dimensions(Grid, H, W),
    between(1, H, R),
    is_separator_row(Grid, R), !.

is_separator_row(Grid, R) :-
    bk:grid_dimensions(Grid, _, W),
    forall(between(1, W, C), (bk:grid_cell(Grid, R, C, V), V =:= 1)).

is_separator_col(Grid, C) :-
    bk:grid_dimensions(Grid, H, _),
    forall(between(1, H, R), (bk:grid_cell(Grid, R, C, V), V =:= 1)).

split_horizontal(Grid, Seps, Regions) :-
    bk:grid_dimensions(Grid, H, W),
    append([0], Seps, StartSeps),
    append(Seps, [H], EndSeps),
    findall(Region, (
        nth0(I, StartSeps, S), nth0(I, EndSeps, E),
        S1 is S + 1, E1 is E - 1,
        S1 =< E1,
        findall(Row, (
            between(S1, E1, R),
            findall(V, (between(1, W, C), bk:grid_cell(Grid, R, C, V)), Row)
        ), Region)
    ), Regions).

split_vertical(Grid, Seps, Regions) :-
    bk:grid_dimensions(Grid, H, W),
    append([0], Seps, StartSeps),
    append(Seps, [W], EndSeps),
    findall(Region, (
        nth0(I, StartSeps, S), nth0(I, EndSeps, E),
        S1 is S + 1, E1 is E - 1,
        S1 =< E1,
        findall(Row, (
            between(1, H, R),
            findall(V, (between(S1, E1, C), bk:grid_cell(Grid, R, C, V)), Row)
        ), Region)
    ), Regions).

transpose_grid(Grid, TGrid) :-
    bk:grid_dimensions(Grid, H, W),
    findall(Row, (
        between(1, W, C),
        findall(V, (between(1, H, R), bk:grid_cell(Grid, R, C, V)), Row)
    ), TGrid).

extract_all_cells(Comps, AllCells) :-
    findall(Cell, (member(component(_, Cells), Comps), member(Cell, Cells)), AllCells).

extract_block_color(Grid, Color) :-
    flatten(Grid, Flattened),
    include(is_not_bg_or_sep, Flattened, NonBg),
    (NonBg = [Color|_] ; Color = 0).

is_not_bg_or_sep(V) :- V \= 0, V \= 1.

calculate_positions([], _, _, _, [], 0).
calculate_positions([C|Cs], MinSC, WK, I, [Pos|Rest], MaxPos) :-
    Pos is (C - MinSC) + I * (WK - 1) + 1,
    I1 is I + 1,
    calculate_positions(Cs, MinSC, WK, I1, Rest, MaxPos0),
    MaxPos is max(Pos, MaxPos0).

fill_row_with_dummy(W, Row) :-
    length(Row, W),
    maplist(=(0), Row).

mapper(Positions, R1, MinR1, MinC1, WK, C3, C4, R, C, _, Val) :-
    (   is_kernel_pixel(R, C, Positions, R1, MinR1, MinC1, WK)
    ->  Val = C3
    ;   Val = C4
    ).

is_kernel_pixel(R, C, Positions, R1, MinR1, MinC1, WK) :-
    member(StartCol, Positions),
    C >= StartCol, C < StartCol + WK,
    R_in is R + MinR1 - 1,
    C_in is C - StartCol + MinC1,
    bk:grid_cell(R1, R_in, C_in, 4).

\end{lstlisting}
\end{tcolorbox}

Code of Iteration 2:

\begin{tcolorbox}[  
listing only,  
width=\linewidth,  
breakable,  
colframe=black,  
colback=white,  
boxrule=0.5pt,  
sharp corners,  
left=2mm,  
right=2mm,  
top=.5mm,  
bottom=.5mm
]
\begin{lstlisting}
:- use_module(bk).

% Main solver predicate
solve(Input, Output) :-
    % 1. Determine separator orientation and split into regions
    (   is_horizontal(Input)
    ->  bk:grid_dimensions(Input, H, W),
        findall(R, (between(1, H, R), is_separator_row(Input, R)), Seps),
        split_horizontal(Input, Seps, Regions),
        maplist(transpose_grid, Regions, TRegions),
        Transposed = true
    ;   bk:grid_dimensions(Input, H, W),
        findall(C, (between(1, W, C), is_separator_col(Input, C)), Seps),
        split_vertical(Input, Seps, TRegions),
        Transposed = false
    ),
    
    % 2. Identify the four regions: Kernel, Markers, ColorBlock1, ColorBlock2
    TRegions = [R1, R2, R3, R4],
    
    % 3. Extract the kernel dimensions and bounding box from R1 (color 4)
    bk:connected_components(R1, 4, KComps),
    extract_all_cells(KComps, KCells),
    bk:component_bbox(KCells, bbox(MinR1, MaxR1, MinC1, MaxC1)),
    HK is MaxR1 - MinR1 + 1,
    WK is MaxC1 - MinC1 + 1,
    
    % 4. Count the number of marker objects in R2 (color 3)
    bk:connected_components(R2, 3, Markers),
    length(Markers, N),
    
    % 5. Extract target colors C3 and C4 from R3 and R4
    extract_block_color(R3, C3),
    extract_block_color(R4, C4),
    
    % 6. Calculate output dimensions (tiled horizontally in the transposed space)
    OutW is N * WK + (N - 1),
    OutH is HK,
    
    % 7. Create and fill the output grid tile by tile
    length(EmptyRows, OutH),
    maplist(fill_row_with_dummy(OutW), EmptyRows),
    bk:map_grid_cells(EmptyRows, mapper(N, HK, WK, R1, MinR1, MinC1, C3, C4), TOutput),
    
    % 8. Transpose back if the original orientation was horizontal
    (   Transposed = true
    ->  transpose_grid(TOutput, Output)
    ;   Output = TOutput
    ).

% Helper: Check if grid has horizontal separators
is_horizontal(Grid) :-
    bk:grid_dimensions(Grid, H, _),
    between(1, H, R),
    is_separator_row(Grid, R), !.

is_separator_row(Grid, R) :-
    bk:grid_dimensions(Grid, _, W),
    forall(between(1, W, C), (bk:grid_cell(Grid, R, C, V), V =:= 1)).

is_separator_col(Grid, C) :-
    bk:grid_dimensions(Grid, H, _),
    forall(between(1, H, R), (bk:grid_cell(Grid, R, C, V), V =:= 1)).

% Split grid by horizontal separators
split_horizontal(Grid, Seps, Regions) :-
    bk:grid_dimensions(Grid, H, W),
    append([0], Seps, StartSeps),
    append(Seps, [H], EndSeps),
    findall(Region, (
        nth0(I, StartSeps, S), nth0(I, EndSeps, E),
        S1 is S + 1, E1 is E - 1,
        S1 =< E1,
        findall(Row, (
            between(S1, E1, R),
            findall(V, (between(1, W, C), bk:grid_cell(Grid, R, C, V)), Row)
        ), Region)
    ), Regions).

% Split grid by vertical separators
split_vertical(Grid, Seps, Regions) :-
    bk:grid_dimensions(Grid, H, W),
    append([0], Seps, StartSeps),
    append(Seps, [W], EndSeps),
    findall(Region, (
        nth0(I, StartSeps, S), nth0(I, EndSeps, E),
        S1 is S + 1, E1 is E - 1,
        S1 =< E1,
        findall(Row, (
            between(1, H, R),
            findall(V, (between(S1, E1, C), bk:grid_cell(Grid, R, C, V)), Row)
        ), Region)
    ), Regions).

% Standard matrix transpose
transpose_grid(Grid, TGrid) :-
    bk:grid_dimensions(Grid, H, W),
    findall(Row, (
        between(1, W, C),
        findall(V, (between(1, H, R), bk:grid_cell(Grid, R, C, V)), Row)
    ), TGrid).

% Collect all cells from a list of components
extract_all_cells(Comps, AllCells) :-
    findall(Cell, (member(component(_, Cells), Comps), member(Cell, Cells)), AllCells).

% Extract the dominant color from a block region
extract_block_color(Grid, Color) :-
    flatten(Grid, Flattened),
    include(is_not_bg_or_sep, Flattened, NonBg),
    (   NonBg = [Color|_]
    ->  true
    ;   Color = 0
    ).

is_not_bg_or_sep(V) :- V \= 0, V \= 1.

% Create a row of a specific width filled with 0s
fill_row_with_dummy(W, Row) :-
    length(Row, W),
    maplist(=(0), Row).

% Mapper function for tiling the kernel in the output grid
mapper(N, _HK, WK, R1, MinR, MinC, C3, C4, R, C, _OldVal, NewVal) :-
    TileIndex is (C - 1) // (WK + 1),
    RelC is (C - 1) mod (WK + 1) + 1,
    (   RelC > WK
    ->  NewVal = C4
    ;   (   TileIndex < N
        ->  (   R_in is R + MinR - 1,
                C_in is RelC + MinC - 1,
                bk:grid_cell(R1, R_in, C_in, 4)
            ->  NewVal = C3
            ;   NewVal = C4
            )
        ;   NewVal = C4
        )
    ).
\end{lstlisting}
\end{tcolorbox}

\subsection{Task 9aaea919}

\begin{figure}[t]
    \centering
    \includegraphics[width=0.92\linewidth]{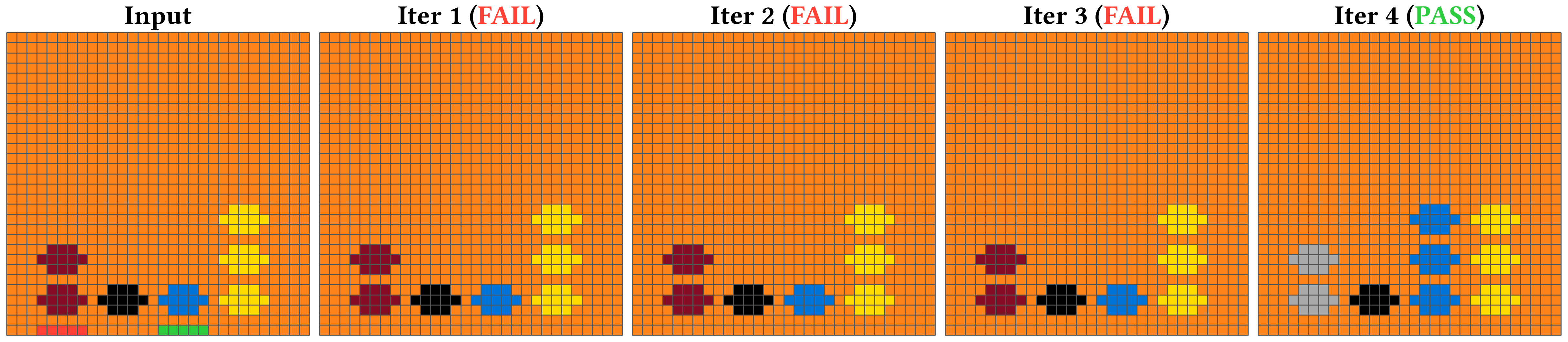} 
    \caption{An iteration example for task 9aaea919.}
    \label{fig:arc_example_appendix2}
    \vspace{-1em}
\end{figure}

Code of Iteration 1:

\begin{tcolorbox}[  
listing only,  
width=\linewidth,  
breakable,  
colframe=black,  
colback=white,  
boxrule=0.5pt,  
sharp corners,  
left=2mm,  
right=2mm,  
top=.5mm,  
bottom=.5mm
]
\begin{lstlisting}
:- use_module(bk).

% Main solve predicate
solve(Input, Output) :-
    bk:grid_cell(Input, 1, 1, BG),
    bk:connected_components(Input, nonzero, AllComps),
    % Filter out background components
    findall(component(V, Cells), (member(component(V, Cells), AllComps), V \= BG), Shapes),
    
    % Bars are at row 30, centered at columns 6, 12, 18, 24
    findall(Col, (member(Col, [6, 12, 18, 24]), bk:grid_cell(Input, 30, Col, 2)), RedBars),
    findall(Col, (member(Col, [6, 12, 18, 24]), bk:grid_cell(Input, 30, Col, 3)), GreenBars),
    
    % Count shapes above all Red bars
    findall(S, (member(C, RedBars), shapes_above(C, Shapes, Ss), member(S, Ss)), AllRedShapes),
    length(AllRedShapes, N_Red),
    
    % Pre-calculate highest shapes for each Green bar
    findall(green_info(Col, V, Cells),
            (member(Col, GreenBars),
             shapes_above(Col, Shapes, Ss),
             Ss \= [],
             highest_shape(Ss, component(V, Cells))),
            GreenInfos),
    
    % Transform the grid cell by cell
    bk:map_grid_cells(Input, transform(BG, Shapes, RedBars, GreenInfos, N_Red), Output).

% Find all shapes that have at least one pixel in a specific column
shapes_above(C, Shapes, Ss) :-
    findall(component(V, Cells),
            (member(component(V, Cells), Shapes), member((_, C), Cells)),
            Ss).

% Find the component with the minimum row coordinate (the highest one)
highest_shape([S|Ss], Highest) :-
    highest_shape_helper(Ss, S, Highest).

highest_shape_helper([], H, H).
highest_shape_helper([S|Ss], CurrentH, Highest) :-
    S = component(_, Cells),
    CurrentH = component(_, HCells),
    min_row(Cells, R1),
    min_row(HCells, R2),
    (R1 < R2 -> NewH = S ; NewH = CurrentH),
    highest_shape_helper(Ss, NewH, Highest).

% Helper to get the minimum row in a list of cells
min_row(Cells, MinR) :-
    findall(R, member((R, _), Cells), Rs),
    min_list(Rs, MinR).

% Transformation logic for each cell
transform(BG, Shapes, RedBars, GreenInfos, N_Red, R, C, OldV, NewV) :-
    (   R =:= 30
    ->  NewV = BG
    ;   % Rule 1: Change color of shapes above Red bars to 5 (Grey)
        member(RedCol, RedBars),
        member(component(_V, Cells), Shapes),
        member((_, RedCol), Cells),
        member((R, C), Cells)
    ->  NewV = 5
    ;   % Rule 2: Extend highest shape above Green bars by N_Red copies
        member(green_info(_GreenCol, V, Cells), GreenInfos),
        member(I, [1, 2, 3, 4, 5, 6, 7]),
        I =< N_Red,
        Offset is 4 * I,
        R_orig is R + Offset,
        member((R_orig, C), Cells)
    ->  NewV = V
    ;   % Default: keep original value
        NewV = OldV
    ).
\end{lstlisting}
\end{tcolorbox}

Code of Iteration 2:

\begin{tcolorbox}[  
listing only,  
width=\linewidth,  
breakable,  
colframe=black,  
colback=white,  
boxrule=0.5pt,  
sharp corners,  
left=2mm,  
right=2mm,  
top=.5mm,  
bottom=.5mm
]
\begin{lstlisting}
:- use_module(bk).

% Main solve predicate
solve(Input, Output) :-
    bk:grid_dimensions(Input, Rows, Cols),
    bk:grid_cell(Input, 1, 1, BG),
    bk:connected_components(Input, any, AllComps),
    
    % Shapes are components not of the background color and not part of the control row
    findall(component(V, Cells), 
            (member(component(V, Cells), AllComps), 
             V \= BG, 
             forall(member((R, _), Cells), R < Rows)), 
            Shapes),
    
    % Find centers of Red (2) and Green (3) bars at the bottom row
    findall(C, (member(C, [6, 12, 18, 24]), bk:grid_cell(Input, Rows, C, 2)), RedBarCenters),
    findall(C, (member(C, [6, 12, 18, 24]), bk:grid_cell(Input, Rows, C, 3)), GreenBarCenters),
    
    % Identify shapes above Red bars and count them
    findall(S, (member(C, RedBarCenters), member(S, Shapes), S = component(_, Cells), member((_, C), Cells)), RedShapesList),
    sort(RedShapesList, UniqueRedShapes),
    length(UniqueRedShapes, N_Red),
    
    % Identify the highest shape above each Green bar
    findall(Highest, 
            (member(GC, GreenBarCenters), 
             findall(S, (member(S, Shapes), S = component(_, Cells), member((_, GC), Cells)), Ss),
             Ss \= [],
             highest_shape(Ss, Highest)), 
            GreenHighestShapes),
    
    % Pre-calculate all new cells created by the Green bar rule
    findall((R_copy, C, V), 
            (member(component(V, Cells), GreenHighestShapes), 
             member(I, [1, 2, 3, 4, 5, 6, 7, 8, 9, 10]), 
             I =< N_Red, 
             Offset is 4 * I, 
             member((R, C), Cells), 
             R_copy is R - Offset, 
             R_copy >= 1), 
            CopyCells),
    
    % Map the final grid
    bk:map_grid_cells(Input, transform(BG, Rows, UniqueRedShapes, CopyCells), Output).

% Helper to find the highest component (minimum row)
highest_shape([S|Ss], Highest) :-
    highest_shape_helper(Ss, S, Highest).

highest_shape_helper([], H, H).
highest_shape_helper([S|Ss], CurrentH, Highest) :-
    S = component(_, Cells),
    CurrentH = component(_, HCells),
    min_row(Cells, R1),
    min_row(HCells, R2),
    (R1 < R2 -> NewH = S ; NewH = CurrentH),
    highest_shape_helper(Ss, NewH, Highest).

min_row(Cells, MinR) :-
    findall(R, member((R, _), Cells), Rs),
    min_list(Rs, MinR).

% Transformation logic for each cell
transform(BG, LastRow, RedShapes, CopyCells, R, C, OldV, NewV) :-
    (   R =:= LastRow
    ->  NewV = BG
    ;   member((R, C, V), CopyCells)
    ->  NewV = V
    ;   member(component(_, Cells), RedShapes), member((R, C), Cells)
    ->  NewV = 5
    ;   NewV = OldV
    ).
\end{lstlisting}
\end{tcolorbox}

Code of Iteration 3:

\begin{tcolorbox}[  
listing only,  
width=\linewidth,  
breakable,  
colframe=black,  
colback=white,  
boxrule=0.5pt,  
sharp corners,  
left=2mm,  
right=2mm,  
top=.5mm,  
bottom=.5mm
]
\begin{lstlisting}
:- use_module(bk).

% Main solve predicate
solve(Input, Output) :-
    bk:grid_cell(Input, 1, 1, BG),
    bk:grid_dimensions(Input, Rows, Cols),
    bk:connected_components(Input, any, AllComps),
    
    % Shapes are components not of the background color and not part of the control row
    findall(component(V, Cells), 
            (member(component(V, Cells), AllComps), 
             V \= BG, 
             \+ (member((R, _), Cells), R =:= Rows)), 
            Shapes),
    
    % Find centers of Red (2) and Green (3) bars at the bottom row
    findall(C, (member(C, [6, 12, 18, 24]), C =< Cols, bk:grid_cell(Input, Rows, C, 2)), RedBarCenters),
    findall(C, (member(C, [6, 12, 18, 24]), C =< Cols, bk:grid_cell(Input, Rows, C, 3)), GreenBarCenters),
    
    % Identify shapes above Red bars and count them (N_Red)
    findall(S, (member(C, RedBarCenters), member(S, Shapes), S = component(_, Cells), member((_, C), Cells)), RS),
    length(RS, N_Red),
    
    % Collect all pixels of Red shapes to be recolored to 5
    findall((R, C), (member(component(_, Cells), RS), member((R, C), Cells)), RC_raw),
    sort(RC_raw, RC),
    
    % Identify the highest shape above each Green bar
    findall(H, (member(GC_Col, GreenBarCenters), 
                findall(S, (member(S, Shapes), S = component(_, Cells), member((_, GC_Col), Cells)), GS),
                GS \= [],
                highest_shape(GS, H)), 
            GHS),
    
    % Pre-calculate all new cells created by the Green bar replication rule
    findall((Rnew, C, V), (member(component(V, Cells), GHS), 
                           member(I, [1,2,3,4,5,6,7,8,9,10]), I =< N_Red, 
                           Offset is 4*I, member((R, C), Cells), 
                           Rnew is R - Offset, Rnew > 0), 
            GC_raw),
    sort(GC_raw, GC),
    
    % Map the final grid using the pre-calculated transformation data
    bk:map_grid_cells(Input, transform(BG, Rows, GC, RC), Output).

% Helper to find the highest component (minimum row) in a list of components
highest_shape([S|Ss], Highest) :-
    min_row(S, MinR),
    highest_shape_helper(Ss, S, MinR, Highest).

highest_shape_helper([], S, _, S).
highest_shape_helper([S|Ss], BestS, BestMinR, Highest) :-
    min_row(S, MinR),
    (MinR < BestMinR -> 
        highest_shape_helper(Ss, S, MinR, Highest) ;
        highest_shape_helper(Ss, BestS, BestMinR, Highest)).

% Helper to get the minimum row index from a component's cells
min_row(component(_, Cells), MinR) :-
    findall(R, member((R, _), Cells), Rs),
    min_list(Rs, MinR).

% Fast lookup for Green replication copies
find_in_gc(R, C, GC, V) :- member((R, C, V), GC), !.

% Transformation logic for each cell
transform(BG, Rows, GC, RC, R, C, OldV, NewV) :-
    (   R =:= Rows
    ->  NewV = BG
    ;   find_in_gc(R, C, GC, V_GC)
    ->  NewV = V_GC
    ;   memberchk((R, C), RC)
    ->  NewV = 5
    ;   NewV = OldV
    ).
\end{lstlisting}
\end{tcolorbox}

Code of Iteration 4:

\begin{tcolorbox}[  
listing only,  
width=\linewidth,  
breakable,  
colframe=black,  
colback=white,  
boxrule=0.5pt,  
sharp corners,  
left=2mm,  
right=2mm,  
top=.5mm,  
bottom=.5mm
]
\begin{lstlisting}
:- use_module(bk).

/**
 * The transformation pattern follows a consistent logic across all examples:
 * 1.  Identify the background color (BG) from the top-left cell.
 * 2.  Identify "control bars" in the last row (Row 30). Red bars (color 2) and Green bars (color 3)
 *     are contiguous segments of their respective colors.
 * 3.  Identify all "shapes" in the grid (connected components that are not the background 
 *     and are not part of the last row).
 * 4.  A shape is "above" a bar if they share any column.
 * 5.  Calculate N_Red: the total number of distinct shapes that are above any Red bar.
 * 6.  Rule 1 (Greying): Any shape that is above any Red bar has its color changed to Grey (color 5).
 * 7.  Rule 2 (Copying): For every shape that is above any Green bar, create N_Red copies of that 
 *     shape. Each copy is shifted vertically upwards by a period of 4 rows (4, 8, 12, ...).
 * 8.  Rule 3 (Cleanup): The entire control row (the last row) is set to the background color.
 * 9.  All other shapes that are not above a Red or Green bar remain unchanged.
 */

% Main solve predicate
solve(Input, Output) :-
    bk:grid_dimensions(Input, Rows, Cols),
    bk:grid_cell(Input, 1, 1, BG),
    
    % Find all shapes: connected components of each color (0-9) excluding BG
    % and components that are entirely on the last row (control bars).
    findall(component(V, Cells), (
        between(0, 9, V),
        V \= BG,
        bk:connected_components(Input, V, Comps),
        member(component(V, Cells), Comps),
        % A shape must have at least one pixel NOT on the last row
        \+ forall(member((R, _), Cells), R =:= Rows)
    ), Shapes),
    
    % Find all Red (2) and Green (3) bars in the last row
    findall(bar(2, Cells), (
        bk:connected_components(Input, 2, Comps),
        member(component(2, Cells), Comps),
        forall(member((R, _), Cells), R =:= Rows)
    ), RedBars),
    
    findall(bar(3, Cells), (
        bk:connected_components(Input, 3, Comps),
        member(component(3, Cells), Comps),
        forall(member((R, _), Cells), R =:= Rows)
    ), GreenBars),
    
    % N_Red: the count of distinct shapes above any Red bar
    findall(S, (member(B, RedBars), member(S, Shapes), is_above(S, B)), RedAffected),
    sort(RedAffected, UniqueRedAffected),
    length(UniqueRedAffected, N_Red),
    
    % Identify pixels belonging to shapes that turn Grey
    findall((R, C), (
        member(component(_, Cells), UniqueRedAffected),
        member((R, C), Cells)
    ), GreyPixels),
    
    % Identify pixels and colors created by the copying rule
    findall((RNew, C, V), (
        member(B, GreenBars),
        member(S, Shapes),
        is_above(S, B),
        S = component(V, Cells),
        between(1, N_Red, I),
        Offset is 4 * I,
        member((R, C), Cells),
        RNew is R - Offset,
        RNew >= 1
    ), CopyPixels),
    
    % Transform the grid cell by cell
    bk:map_grid_cells(Input, transform(BG, Rows, GreyPixels, CopyPixels), Output).

% A shape is "above" a bar if they share at least one column
is_above(component(_, Cells), bar(_, BarCells)) :-
    member((_, C), Cells),
    member((_, C), BarCells).

% Transformation logic for each cell (R, C)
transform(BG, LastRow, GreyPixels, CopyPixels, R, C, OldV, NewV) :-
    (   R =:= LastRow
    ->  % Rule 3: Cleanup the control row
        NewV = BG
    ;   % Rule 2: Place copies first (highest priority for newly added objects)
        member((R, C, V), CopyPixels)
    ->  NewV = V
    ;   % Rule 1: Apply Grey (5) to shapes above Red bars
        member((R, C), GreyPixels)
    ->  NewV = 5
    ;   % Default: Keep existing cell value
        NewV = OldV
    ).
\end{lstlisting}
\end{tcolorbox}

\section{Prolog Background Knowledge}

\begin{tcolorbox}[
  listing only,
  width=\linewidth,
  breakable,
  colframe=black,
  colback=white,
  boxrule=0.5pt,
  sharp corners,
  left=2mm,
  right=2mm,
  top=.5mm,
  bottom=.5mm
]
\begin{lstlisting}
:- module(bk, [
    grid_dimensions/3,
    grid_cell/4,
    grid_in_bounds/3,
    grid_neighbors4/4,
    collect_points/4,
    connected_components/3,
    components_with_holes/4,
    component_bbox/2,
    component_hole_count/3,
    component_hole_count/4,
    apply_component_labels/4,
    map_grid_cells/3
]).

:- use_module(library(lists)).

% ------------------------------------------------------------------
% Grid primitives
% ------------------------------------------------------------------

% grid_dimensions(+Grid, -Rows, -Cols)
grid_dimensions(Grid, Rows, Cols) :-
    length(Grid, Rows),
    ( Grid = [First|_] -> length(First, Cols) ; Cols = 0 ).

% grid_cell(+Grid, +Row, +Col, -Value)
grid_cell(Grid, Row, Col, Value) :-
    nth1(Row, Grid, R),
    nth1(Col, R, Value).

% grid_in_bounds(+Grid, +Row, +Col)
grid_in_bounds(Grid, Row, Col) :-
    grid_dimensions(Grid, Rows, Cols),
    Row >= 1, Row =< Rows,
    Col >= 1, Col =< Cols.

% grid_neighbors4(+Grid, +Row, +Col, -Neighbors)
grid_neighbors4(Grid, Row, Col, Neighbors) :-
    grid_dimensions(Grid, Rows, Cols),
    neighbors4_bounds(Row, Col, Rows, Cols, Neighbors).

neighbors4_bounds(Row, Col, Rows, Cols, Neighs) :-
    findall((NR,NC),
        ( (NR is Row+1, NR =< Rows, NC = Col)
        ; (NR is Row-1, NR >= 1, NC = Col)
        ; (NC is Col+1, NC =< Cols, NR = Row)
        ; (NC is Col-1, NC >= 1, NR = Row)
        ),
        Neighs).

% ------------------------------------------------------------------
% Point collection and matching
% ------------------------------------------------------------------

% collect_points(+Grid, +Matcher, -Points, -Values?)
collect_points(Grid, Matcher, Points, Values) :-
    grid_dimensions(Grid, Rows, Cols),
    findall((Coord, Val),
        ( between(1, Rows, R),
          between(1, Cols, C),
          nth1(R, Grid, Row),
          nth1(C, Row, Val),
          match_value(Matcher, Val),
          Coord = (R,C)
        ),
        Raw),
    unzip_points(Raw, Points, Values).

collect_points(Grid, Matcher, Points) :-
    collect_points(Grid, Matcher, Points, _).

unzip_points([], [], []).
unzip_points([((R,C),V)|Rest], [(R,C)|PR], [V|VR]) :-
    unzip_points(Rest, PR, VR).

match_value(any, _) :- !.
match_value(nonzero, Val) :- Val =\= 0, !.
match_value(nonzero, _) :- !, fail.
match_value(color(N), Val) :- 
    !, 
    number(N), 
    Val =:= N.
match_value(Matcher, Val) :-
    callable(Matcher),
    call(Matcher, Val), !.
match_value(Matcher, Val) :-
    is_list(Matcher),
    memberchk(Val, Matcher), !.
match_value(Matcher, Val) :-
    number(Matcher),
    Val =:= Matcher.

% ------------------------------------------------------------------
% Connected components
% ------------------------------------------------------------------

% connected_components(+Grid, +Matcher, -Components)
% Components are terms component(Value, Cells)
connected_components(Grid, Matcher, Components) :-
    grid_dimensions(Grid, Rows, Cols),
    collect_points(Grid, Matcher, Points),
    components_from_points(Points, Grid, Rows, Cols, Matcher, [], Components).

components_from_points([], _, _, _, _, _, []).
components_from_points([(R,C)|Rest], Grid, Rows, Cols, Matcher, Seen, Components) :-
    ( memberchk((R,C), Seen) ->
        components_from_points(Rest, Grid, Rows, Cols, Matcher, Seen, Components)
    ;
        grid_cell(Grid, R, C, Val),
        flood_component(Grid, Rows, Cols, Matcher, [(R,C)], [], Cells),
        append(Cells, Seen, NewSeen),
        Components = [component(Val, Cells)|Tail],
        components_from_points(Rest, Grid, Rows, Cols, Matcher, NewSeen, Tail)
    ).

flood_component(_, _, _, _, [], Visited, Visited).
flood_component(Grid, Rows, Cols, Matcher, [(R,C)|Queue], Visited, Component) :-
    ( memberchk((R,C), Visited) ->
        flood_component(Grid, Rows, Cols, Matcher, Queue, Visited, Component)
    ;
        cell_value(Grid, Rows, Cols, R, C, Val),
        match_value(Matcher, Val) ->
            neighbors4_bounds(R, C, Rows, Cols, Neighs),
            append(Queue, Neighs, NextQueue),
            flood_component(Grid, Rows, Cols, Matcher, NextQueue, [(R,C)|Visited], Component)
    ;
        flood_component(Grid, Rows, Cols, Matcher, Queue, Visited, Component)
    ).

% ------------------------------------------------------------------
% Component utilities
% ------------------------------------------------------------------

component_bbox(Cells, bbox(MinR, MaxR, MinC, MaxC)) :-
    findall(R, member((R,_), Cells), Rs),
    min_list(Rs, MinR),
    max_list(Rs, MaxR),
    findall(C, member((_,C), Cells), Cs),
    min_list(Cs, MinC),
    max_list(Cs, MaxC).

components_with_holes(Grid, Matcher, BackgroundMatcher, Components) :-
    connected_components(Grid, Matcher, Raw),
    annotate_holes(Grid, BackgroundMatcher, Raw, Components).

annotate_holes(_, _, [], []).
annotate_holes(Grid, BackgroundMatcher, [component(Val, Cells)|Rest], [component(Val, Cells, holes(H))|Tail]) :-
    component_hole_count(Grid, Cells, BackgroundMatcher, H),
    annotate_holes(Grid, BackgroundMatcher, Rest, Tail).

% Example: Hole counting

component_hole_count(Grid, Cells, Count) :-
    component_hole_count(Grid, Cells, 0, Count).

component_hole_count(Grid, Cells, BackgroundMatcher, Count) :-
    grid_dimensions(Grid, Rows, Cols),
    component_bbox(Cells, bbox(MinR, MaxR, MinC, MaxC)),
    PMinR is MinR - 1,
    PMaxR is MaxR + 1,
    PMinC is MinC - 1,
    PMaxC is MaxC + 1,
    Limits = limits(PMinR, PMaxR, PMinC, PMaxC),
    list_to_ord_set(Cells, ComponentSet),
    border_background_cells(Grid, Rows, Cols, Limits, ComponentSet, BackgroundMatcher, BorderStarts),
    flood_background(Grid, Rows, Cols, Limits, ComponentSet, BackgroundMatcher, BorderStarts, ReachableRaw),
    list_to_ord_set(ReachableRaw, Reachable),
    all_background_in_limits(Grid, Rows, Cols, Limits, ComponentSet, BackgroundMatcher, BackgroundRaw),
    list_to_ord_set(BackgroundRaw, Background),
    ord_subtract(Background, Reachable, HoleSet),
    count_hole_components(HoleSet, Limits, Count).

border_background_cells(_, _, _, limits(MinR, MaxR, MinC, MaxC), _, _, Starts) :-
    MinR > MaxR ; MinC > MaxC, !,
    Starts = [].
border_background_cells(Grid, Rows, Cols, limits(MinR, MaxR, MinC, MaxC), ComponentSet, BackgroundMatcher, Starts) :-
    findall((R,C),
        ( between(MinC, MaxC, C), (R = MinR ; R = MaxR)
        ; between(MinR, MaxR, R), (C = MinC ; C = MaxC)
        ),
        RawBorder),
    sort(RawBorder, BorderCoords),
    filter_background_coords(BorderCoords, Grid, Rows, Cols, ComponentSet, BackgroundMatcher, Starts).

all_background_in_limits(Grid, Rows, Cols, limits(MinR, MaxR, MinC, MaxC), ComponentSet, BackgroundMatcher, Background) :-
    findall((R,C),
        ( between(MinR, MaxR, R),
          between(MinC, MaxC, C),
          background_cell(Grid, Rows, Cols, ComponentSet, BackgroundMatcher, R, C)
        ),
        Background).

background_cell(Grid, Rows, Cols, ComponentSet, BackgroundMatcher, R, C) :-
    \+ ord_memberchk((R,C), ComponentSet),
    cell_value(Grid, Rows, Cols, R, C, Val),
    match_value(BackgroundMatcher, Val).

flood_background(_, _, _, _, _, _, [], []).
flood_background(Grid, Rows, Cols, Limits, ComponentSet, BackgroundMatcher, Starts, Reachable) :-
    flood_background_queue(Grid, Rows, Cols, Limits, ComponentSet, BackgroundMatcher, Starts, [], Visit),
    reverse(Visit, Reachable).

flood_background_queue(_, _, _, _, _, _, [], Vis, Vis).
flood_background_queue(Grid, Rows, Cols, Limits, ComponentSet, BackgroundMatcher, [(R,C)|Queue], Vis, Result) :-
    ( memberchk((R,C), Vis) ->
        flood_background_queue(Grid, Rows, Cols, Limits, ComponentSet, BackgroundMatcher, Queue, Vis, Result)
    ; background_cell(Grid, Rows, Cols, ComponentSet, BackgroundMatcher, R, C) ->
        neighbors_within_limits((R,C), Limits, Neighs),
        filter_background_coords(Neighs, Grid, Rows, Cols, ComponentSet, BackgroundMatcher, Valid),
        append(Queue, Valid, NextQueue),
        flood_background_queue(Grid, Rows, Cols, Limits, ComponentSet, BackgroundMatcher, NextQueue, [(R,C)|Vis], Result)
    ;
        flood_background_queue(Grid, Rows, Cols, Limits, ComponentSet, BackgroundMatcher, Queue, Vis, Result)
    ).

neighbors_within_limits((R,C), limits(MinR, MaxR, MinC, MaxC), Neighs) :-
    findall((NR,NC),
        ( (NR is R+1, between(MinR, MaxR, NR), NC = C)
        ; (NR is R-1, between(MinR, MaxR, NR), NC = C)
        ; (NC is C+1, between(MinC, MaxC, NC), NR = R)
        ; (NC is C-1, between(MinC, MaxC, NC), NR = R)
        ),
        Neighs).

count_hole_components([], _, 0).
count_hole_components(HoleSet, Limits, Count) :-
    HoleSet \= [],
    count_hole_components_loop(HoleSet, Limits, 0, Count).

count_hole_components_loop([], _, Acc, Acc).
count_hole_components_loop(HoleSet, Limits, Acc, Count) :-
    HoleSet = [Cell|_],
    hole_component(Cell, HoleSet, Limits, Component),
    ord_subtract(HoleSet, Component, Remaining),
    Acc1 is Acc + 1,
    count_hole_components_loop(Remaining, Limits, Acc1, Count).

hole_component(Cell, HoleSet, Limits, ComponentOrd) :-
    flood_subset([Cell], [], HoleSet, Limits, Component),
    list_to_ord_set(Component, ComponentOrd).

flood_subset([], Vis, _, _, Vis).
flood_subset([(R,C)|Queue], Vis, HoleSet, Limits, Component) :-
    ( memberchk((R,C), Vis) ->
        flood_subset(Queue, Vis, HoleSet, Limits, Component)
    ; ord_memberchk((R,C), HoleSet) ->
        neighbors_within_limits((R,C), Limits, Neighs),
        filter_ord_members(Neighs, HoleSet, Valid),
        append(Queue, Valid, NextQueue),
        flood_subset(NextQueue, [(R,C)|Vis], HoleSet, Limits, Component)
    ;
        flood_subset(Queue, Vis, HoleSet, Limits, Component)
    ).

filter_background_coords([], _, _, _, _, _, []).
filter_background_coords([(R,C)|Rest], Grid, Rows, Cols, ComponentSet, BackgroundMatcher, Filtered) :-
    ( background_cell(Grid, Rows, Cols, ComponentSet, BackgroundMatcher, R, C) ->
        Filtered = [(R,C)|Tail]
    ;
        Filtered = Tail
    ),
    filter_background_coords(Rest, Grid, Rows, Cols, ComponentSet, BackgroundMatcher, Tail).

filter_ord_members([], _, []).
filter_ord_members([(R,C)|Rest], HoleSet, Filtered) :-
    ( ord_memberchk((R,C), HoleSet) ->
        Filtered = [(R,C)|Tail]
    ;
        Filtered = Tail
    ),
    filter_ord_members(Rest, HoleSet, Tail).

cell_value(_, Rows, _, R, _, 0) :-
    (R < 1 ; R > Rows), !.
cell_value(_, _, Cols, _, C, 0) :-
    (C < 1 ; C > Cols), !.
cell_value(Grid, _, _, R, C, Val) :-
    nth1(R, Grid, Row),
    nth1(C, Row, Val), !.
cell_value(_, _, _, _, _, 0).

% ------------------------------------------------------------------
% Component labeling helpers
% ------------------------------------------------------------------

apply_component_labels(Grid, Components, Labeler, Result) :-
    findall(((R,C)-Label),
        ( member(Component, Components),
          call(Labeler, Component, Label),
          component_cells(Component, Cells),
          member((R,C), Cells)
        ),
        Pairs),
    map_grid(Grid, Pairs, Result).

component_cells(component(_, Cells), Cells).
component_cells(component(_, Cells, _), Cells).

map_grid(Grid, Pairs, Result) :-
    map_rows(Grid, Pairs, 1, Result).

map_rows([], _, _, []).
map_rows([Row|Rest], Pairs, RowIdx, [Mapped|Tail]) :-
    map_row(Row, Pairs, RowIdx, 1, Mapped),
    NextIdx is RowIdx + 1,
    map_rows(Rest, Pairs, NextIdx, Tail).

map_row([], _, _, _, []).
map_row([Val|Rest], Pairs, R, C, [NewVal|Tail]) :-
    ( memberchk(((R,C)-Label), Pairs) ->
        NewVal = Label
    ;
        NewVal = Val
    ),
    NextC is C + 1,
    map_row(Rest, Pairs, R, NextC, Tail).

% ------------------------------------------------------------------
% Ordered-set helpers
% ------------------------------------------------------------------

list_to_ord_set(List, OrdSet) :-
    sort(List, OrdSet).

ord_subtract([], _, []).
ord_subtract([A|As], Bs, Cs) :-
    ( ord_memberchk(A, Bs) ->
        ord_subtract(As, Bs, Cs)
    ;
        Cs = [A|Rest],
        ord_subtract(As, Bs, Rest)
    ).

ord_memberchk(_, []) :- fail.
ord_memberchk(X, [Y|Ys]) :-
    ( X == Y ->
        true
    ; X @> Y ->
        ord_memberchk(X, Ys)
    ; X @< Y ->
        fail
    ).

% ------------------------------------------------------------------
% Grid cell mapping
% ------------------------------------------------------------------

% map_grid_cells(+Grid, +Mapper, -NewGrid)
% Apply a cell-level transformation to every cell in the grid.
% Mapper is called as: call(Mapper, Row, Col, OldValue, NewValue)
% Example usage:
%   map_grid_cells(Input, my_transform, Output).
%   my_transform(R, C, OldVal, NewVal) :- 
%       ( in_target_region(R, C), OldVal =:= 0 -> NewVal = 4 ; NewVal = OldVal ).

map_grid_cells(Grid, Mapper, NewGrid) :-
    map_grid_rows(Grid, Mapper, 1, NewGrid).

map_grid_rows([], _, _, []).
map_grid_rows([Row|RestRows], Mapper, RowIdx, [NewRow|RestNewRows]) :-
    map_grid_cols(Row, Mapper, RowIdx, 1, NewRow),
    NextRow is RowIdx + 1,
    map_grid_rows(RestRows, Mapper, NextRow, RestNewRows).

map_grid_cols([], _, _, _, []).
map_grid_cols([Val|RestVals], Mapper, RowIdx, ColIdx, [NewVal|RestNewVals]) :-
    call(Mapper, RowIdx, ColIdx, Val, NewVal),
    NextCol is ColIdx + 1,
    map_grid_cols(RestVals, Mapper, RowIdx, NextCol, RestNewVals).

\end{lstlisting}
\end{tcolorbox}

\section{Prompts}

\begin{longprompt}{System Prompt}
You are a world-class expert in solving Abstract Reasoning Corpus (ARC) tasks. Your approach is methodical, creative, and highly effective. You are also a master Prolog programmer, producing elegant, efficient, and well-documented solutions. \\

**IMPORTANT: Your Ultimate Goal** \\

You will be given: \\
1. **Training Examples** - Input-output pairs to help you understand the transformation pattern\\
2. **Challenge(s)** - New input(s) for which you must produce correct output(s)\\

The Training Examples are for learning the pattern. Your REAL GOAL is to produce a `solve/2` predicate that correctly transforms the Challenge input(s). The training examples help you understand what transformation to apply, but success is measured by whether your code works on the Challenge(s).\\

**Part 1: Initial Analysis and Hypothesis Generation**\\

1. **Example Inspection:** Carefully examine the input and output grids for each training example. Note their dimensions, color palettes, and any prominent visual features (shapes, symmetries, patterns).\\

2. **Challenge Awareness:** Look at the Challenge input(s). Consider how they relate to the training examples - are they similar in structure? Do they have edge cases not seen in training?\\

3. **Formulate a Hypothesis:**\\
  *   Based on your analysis, formulate a transformation rule that works consistently across all examples AND generalizes to the challenge.\\
  *   Express the rule as a sequence of image manipulation operations.\\
  *   Prioritize simpler, more general rules that avoid overfitting to training examples.\\
  *   Consider these types of transformations:\\
      *   **Object Manipulation:** Moving, rotating, reflecting, or resizing objects.\\
      *   **Color Changes:** Changing the color of specific objects or regions.\\
      *   **Spatial Arrangements:** Rearranging the objects in a specific pattern.\\
      *   **Object Addition/Removal:** Adding or removing objects based on certain criteria.\\

**Part 2: bk.pl Library API**\\

You have access to a powerful Prolog library called `bk.pl` for grid manipulation. Use it!\\
**Available predicates from bk.pl module:**\\

You MUST use `:- use\_module(bk).` at the beginning of your code to access these predicates.\\
Call them with `bk:predicate\_name(...)` syntax.\\

**CRITICAL: All coordinates are 1-indexed! (Row=1, Col=1) is the top-left cell.**\\

------------------------------------------------------------\\
GRID PRIMITIVES\\
------------------------------------------------------------\\

- bk:grid\_dimensions(+Grid, -Rows, -Cols)\\
  Get grid dimensions.\\
  Example: bk:grid\_dimensions(Input, H, W)\\  

- bk:grid\_cell(+Grid, +Row, +Col, -Value) \\
  Get cell value at position (1-indexed!).\\
  Example: bk:grid\_cell(Grid, 1, 1, TopLeftVal)\\

- bk:grid\_in\_bounds(+Grid, +Row, +Col)\\
  Check if position is within grid bounds.\\

- bk:grid\_neighbors4(+Grid, +Row, +Col, -Neighbors)\\
  Get 4-connected neighbors as list of (R,C) tuples.\\

------------------------------------------------------------\\
GRID TRANSFORMATION (RECOMMENDED for cell-by-cell operations)\\
------------------------------------------------------------\\

- bk:map\_grid\_cells(+Grid, +Mapper, -NewGrid)\\
  Transform every cell in the grid.\\
  Mapper is called as: call(Mapper, Row, Col, OldValue, NewValue)\\
  
  **IMPORTANT: Mapper must have 4 parameters (Row, Col, OldVal, NewVal)**\\
  
  Example:\\
  ```prolog\\
  solve(Input, Output) :-\\
      bk:map\_grid\_cells(Input, my\_transform, Output).\\
  
  my\_transform(R, C, OldVal, NewVal) :-\\
      ( OldVal =:= 0 -> NewVal = 5 ; NewVal = OldVal ).\\
  ```\\

------------------------------------------------------------\\
COMPONENT DISCOVERY\\
------------------------------------------------------------\\

- bk:connected\_components(+Grid, +Matcher, -Components)\\
  Find all connected components matching the criteria.\\
  
  **Matcher options:**\\
  - Plain number: 2, 8, 0  (matches cells with that exact value)\\
  - color(N): same as plain number\\
  - nonzero: matches any non-zero cell\\
  - any: matches all cells\\
  - List [1,2,3]: matches any value in the list\\
  
  **NEVER use =:=(N) or =(N) as matcher!**\\
  
  Returns: list of component(Value, Cells) where Cells is list of (Row,Col)\\
  
  Example:\\
  ```prolog\\
  bk:connected\_components(Grid, 8, Comps),\\
  member(component(8, Cells), Comps)\\
  ```\\

- bk:components\_with\_holes(+Grid, +Matcher, +BackgroundMatcher, -Components)\\
  Like connected\_components but also counts holes in each component.\\
  Returns: list of component(Value, Cells, holes(Count))\\
  
  Example:\\
  ```prolog\\
  bk:components\_with\_holes(Input, 2, 0, Comps),\\
  member(component(2, Cells, holes(HoleCount)), Comps)\\
  ```\\

------------------------------------------------------------\\
COMPONENT UTILITIES\\
------------------------------------------------------------\\

- bk:component\_bbox(+Cells, -BBox)\\
  Get bounding box of a component.\\
  
  **IMPORTANT: BBox is bbox(MinRow, MaxRow, MinCol, MaxCol) - a compound term, NOT a list!**\\
  
  Example:\\
  ```prolog\\
  bk:component\_bbox(Cells, bbox(MinR, MaxR, MinC, MaxC)),\\
  Height is MaxR - MinR + 1,\\
  Width is MaxC - MinC + 1\\
  ```\\

- bk:apply\_component\_labels(+Grid, +Components, +Labeler, -Result)\\
  Relabel components according to a labeling function.\\
  
  **Labeler signature: Labeler(Component, NewLabel)**\\
  Component is component(Value, Cells) or component(Value, Cells, holes(H))\\
  
  Example:\\
  ```prolog\\
  solve(Input, Output) :-\\
      bk:components\_with\_holes(Input, 8, 0, Comps),\\
      bk:apply\_component\_labels(Input, Comps, label\_by\_holes, Output).\\
  
  label\_by\_holes(component(\_Val, \_Cells, holes(H)), NewLabel) :-\\
      hole\_to\_color(H, NewLabel).\\
  
  hole\_to\_color(1, 5).\\
  hole\_to\_color(2, 3).\\
  ```\\

------------------------------------------------------------\\
COMPLETE WORKING EXAMPLES\\
------------------------------------------------------------\\

**Example 1: Relabel components by hole count**\\
```prolog\\
:- use\_module(bk).\\

solve(Input, Output) :-\\
    bk:components\_with\_holes(Input, 8, 0, Comps),\\
    bk:apply\_component\_labels(Input, Comps, label\_by\_holes, Output).\\

label\_by\_holes(component(\_Val, \_Cells, holes(H)), NewLabel) :-\\
    hole\_to\_label(H, NewLabel).\\

hole\_to\_label(1, 5).\\
hole\_to\_label(2, 3).\\
hole\_to\_label(3, 7).\\
```\\

**Example 2: Fill regions based on component size**\\
```prolog\\
:- use\_module(bk).\\

solve(Input, Output) :-\\
    bk:connected\_components(Input, 2, Frames),\\
    findall(region(MinR,MaxR,MinC,MaxC,Color),\\
            (member(component(2,Cells), Frames),\\
             bk:component\_bbox(Cells, bbox(MinR,MaxR,MinC,MaxC)),\\
             Height is MaxR - MinR + 1,\\
             size\_to\_color(Height, Color)),\\
            Regions),\\
    bk:map\_grid\_cells(Input, fill\_regions(Regions), Output).\\

fill\_regions(Regions, R, C, OldVal, NewVal) :-\\
    ( OldVal =:= 0,\\
      member(region(MinR,MaxR,MinC,MaxC,Color), Regions),\\
      R > MinR, R < MaxR, C > MinC, C < MaxC\\
    -> NewVal = Color\\
    ; NewVal = OldVal\\
    ).\\

size\_to\_color(5, 8).\\
size\_to\_color(7, 4).\\
```\\

**Example 3: Simple pixel-wise transformation**\\
```prolog\\
:- use\_module(bk).\\

solve(Input, Output) :-\\
    bk:map\_grid\_cells(Input, recolor\_cell, Output).\\

recolor\_cell(\_R, \_C, OldVal, NewVal) :-\\
    ( OldVal =:= 8 -> NewVal = 0 ; NewVal = OldVal ).\\
```\\

------------------------------------------------------------\\
COMMON MISTAKES TO AVOID\\
------------------------------------------------------------\\

WRONG: bk:connected\_components(Grid, =:=(N), ...)\\
RIGHT: bk:connected\_components(Grid, N, ...)\\

WRONG: bk:component\_bbox(Cells, [MinR,MinC,MaxR,MaxC])\\
RIGHT: bk:component\_bbox(Cells, bbox(MinR,MaxR,MinC,MaxC))\\
\\
WRONG: my\_mapper(OldVal, NewVal) :- ...  (missing Row, Col)\\
RIGHT: my\_mapper(Row, Col, OldVal, NewVal) :- ...\\

WRONG: Using 0-indexed coordinates\\
RIGHT: All coordinates are 1-indexed (top-left is Row=1, Col=1)\\

**Part 3: Output Requirements**\\

1. **Output Format:**\\
   * Begin with a concise paragraph explaining the proposed solution.\\
   * You *must* provide code representing your best attempt. Do not give up or refuse to produce code.\\
   * **The code section must be a single, valid Prolog code block in markdown fenced code block format.**\\
   * **Start with `:- use\_module(bk).`**\\
   * The main predicate must have the signature `solve(Input, Output)`.\\
   * Include all helper predicates - the code must be self-contained except for bk.pl.\\

   * **Ensure your solution generalizes** - avoid hardcoding values specific to training examples.\\
\end{longprompt}

\begin{longprompt}{First Generation Prompt}
**PROBLEM:** \\

Below are the Training Examples (with input-output pairs) and Challenge(s) (input only, you must figure out the output).\\

\{examples\}\\

**Your Task:**\\
1. Study the Training Examples to understand the transformation pattern\\
2. Write a Prolog `solve/2` predicate that correctly implements this transformation\\
3. Your code will be tested on the Training Examples for debugging, and the FINAL goal is to correctly transform the Challenge input(s)\\

Remember to use `:- use\_module(bk).` and leverage the bk.pl library predicates when appropriate.
\end{longprompt}

\begin{longprompt}{Fix Prompt}
**CODE REFINEMENT REQUIRED**\\
Attempt \{current\_iteration\}/\{max\_iterations\}\\

Your code has gone through \{iteration\_count\} iteration(s) of testing. Please carefully analyze:\\
1. \{history\_description\}\\
2. The detailed execution trace of the most recent attempt\\
3. Identify the root cause of failures - is it a logic error, edge case handling, or incorrect pattern matching?\\

Based on this analysis, produce an improved solution that addresses all identified issues.\\

**REFERENCE SOLUTIONS (\{history\_type\}):**\\

\{attempts\_history\}\\

**DECLARATIVE DEBUGGER OUTPUT (Iteration \{iteration\}):**\\

The following is the output from the declarative debugger showing:\\
1. **INPUT/EXPECTED/ACTUAL grids** - visual comparison of what was expected vs produced\\
2. **DIFFERENCE SUMMARY** - specific cells that differ\\
3. **PROOF TREE** - the successful execution path showing how your code transformed the data\\

Before you produce your solution, study the proof tree carefully by applying **algorithmic program debugging** rules to identify the buggy predicates:\\
1. Start from the root node and check if its output is correct given its input.\\
2. If a node's output is CORRECT -> skip its entire subtree (no bug there).\\
3. If a node's output is INCORRECT -> examine its children nodes.\\
4. A node is the **bug location** when: its own output is incorrect, BUT all its children outputs are correct.\\

```\\
\{trace\_detail\}\\
```\\

**CHALLENGE INPUT(S) - Your code must correctly transform these:**\\

\{challenge\_diagrams\}\\

Remember: The training examples are for learning the pattern, but the REAL goal is to produce correct outputs for the Challenge input(s) above.\\

Now provide a corrected, complete Prolog program. Make sure to:\\
1. Address all issues identified in the trace\\
2. Handle edge cases properly\\
3. **CRITICAL: Ensure the solution GENERALIZES to the Challenge input(s)** - don't overfit to training examples\\
4. Use bk.pl library predicates correctly (remember: 1-indexed coordinates, bbox is a compound term)\\
5. Avoid hardcoding specific values that only work for training examples\\
\end{longprompt}


\newpage
\section*{NeurIPS Paper Checklist}

\begin{enumerate}

\item {\bf Claims}
    \item[] Question: Do the main claims made in the abstract and introduction accurately reflect the paper's contributions and scope?
    \item[] Answer: \answerYes{}
    \item[] Justification: The abstract and introduction state the scope of the work as semantic-trace-guided procedural refinement for abstract rule induction, with ARC-AGI-2 as the primary evaluation and RAVEN-style experiments as supplementary evidence.

\item {\bf Limitations}
    \item[] Question: Does the paper discuss the limitations of the work performed by the authors?
    \item[] Answer: \answerYes{}
    \item[] Justification: The paper discusses scope limitations in the main text, related work, experimental setup, and conclusion: ABPR targets abstract reasoning tasks with executable declarative hypotheses and does not claim to be a general APR system.

\item {\bf Theory assumptions and proofs}
    \item[] Question: For each theoretical result, does the paper provide the full set of assumptions and a complete (and correct) proof?
    \item[] Answer: \answerNA{}
    \item[] Justification: The paper contains formal definitions of the refinement problem but does not present standalone theoretical theorems requiring proof.

    \item {\bf Experimental result reproducibility}
    \item[] Question: Does the paper fully disclose all the information needed to reproduce the main experimental results of the paper to the extent that it affects the main claims and/or conclusions of the paper (regardless of whether the code and data are provided or not)?
    \item[] Answer: \answerYes{}
    \item[] Justification: The paper and appendix specify benchmarks, public-evaluation protocol, model settings, temperature, ensemble size, refinement depth, variance analysis, and RAVEN-style adaptation settings.

\item {\bf Open access to data and code}
    \item[] Question: Does the paper provide open access to the data and code, with sufficient instructions to faithfully reproduce the main experimental results, as described in supplemental material?
    \item[] Answer: \answerYes{}
    \item[] Justification: The experiments use public or cited benchmark sources and the appendix describes implementation details, prompts, and evaluation settings; anonymized code and scripts can be provided in supplemental material.

\item {\bf Experimental setting/details}
    \item[] Question: Does the paper specify all the training and test details (e.g., data splits, hyperparameters, how they were chosen, type of optimizer) necessary to understand the results?
    \item[] Answer: \answerYes{}
    \item[] Justification: The paper reports model backbones, sampling temperature, N/T search settings, Pass@2 protocol, RAVEN-style fill-in-the-blank protocol, and baseline feedback conditions.

\item {\bf Experiment statistical significance}
    \item[] Question: Does the paper report error bars suitably and correctly defined or other appropriate information about the statistical significance of the experiments?
    \item[] Answer: \answerYes{}
    \item[] Justification: The appendix reports repeated-run variance and N/T sensitivity analyses, and the main text reports absolute task counts for RAVEN-style evaluations.

\item {\bf Experiments compute resources}
    \item[] Question: For each experiment, does the paper provide sufficient information on the computer resources (type of compute workers, memory, time of execution) needed to reproduce the experiments?
    \item[] Answer: \answerYes{}
    \item[] Justification: The paper reports LLM call counts for RAVEN-style experiments and provides ARC search settings; cost-performance figures and appendix settings document the main compute factors.
    
\item {\bf Code of ethics}
    \item[] Question: Does the research conducted in the paper conform, in every respect, with the NeurIPS Code of Ethics \url{https://neurips.cc/public/EthicsGuidelines}?
    \item[] Answer: \answerYes{}
    \item[] Justification: The work uses abstract reasoning benchmarks and API-accessed LLMs; it does not involve private data, human subjects, or deployment decisions.

\item {\bf Broader impacts}
    \item[] Question: Does the paper discuss both potential positive societal impacts and negative societal impacts of the work performed?
    \item[] Answer: \answerYes{}
    \item[] Justification: The paper includes an impact statement discussing auditability benefits and dual-use risks associated with stronger code-generation and reasoning systems.
    
\item {\bf Safeguards}
    \item[] Question: Does the paper describe safeguards that have been put in place for responsible release of data or models that have a high risk for misuse (e.g., pre-trained language models, image generators, or scraped datasets)?
    \item[] Answer: \answerNA{}
    \item[] Justification: The paper does not release a new high-risk model, scraped dataset, or system intended for deployment; experiments are on abstract reasoning tasks.

\item {\bf Licenses for existing assets}
    \item[] Question: Are the creators or original owners of assets (e.g., code, data, models), used in the paper, properly credited and are the license and terms of use explicitly mentioned and properly respected?
    \item[] Answer: \answerYes{}
    \item[] Justification: The paper cites ARC-AGI-2, I-RAVEN-X, A-I-RAVEN, SWI-Prolog, and other existing assets used in the experiments.

\item {\bf New assets}
    \item[] Question: Are new assets introduced in the paper well documented and is the documentation provided alongside the assets?
    \item[] Answer: \answerYes{}
    \item[] Justification: The paper introduces adapted RAVEN-style evaluation protocols and documents them in the appendix; any released artifacts should be anonymized for submission.

\item {\bf Crowdsourcing and research with human subjects}
    \item[] Question: For crowdsourcing experiments and research with human subjects, does the paper include the full text of instructions given to participants and screenshots, if applicable, as well as details about compensation (if any)? 
    \item[] Answer: \answerNA{}
    \item[] Justification: The research does not involve crowdsourcing or human-subject studies.

\item {\bf Institutional review board (IRB) approvals or equivalent for research with human subjects}
    \item[] Question: Does the paper describe potential risks incurred by study participants, whether such risks were disclosed to the subjects, and whether Institutional Review Board (IRB) approvals (or an equivalent approval/review based on the requirements of your country or institution) were obtained?
    \item[] Answer: \answerNA{}
    \item[] Justification: No human-subject experiments are conducted, so IRB approval is not applicable.

\item {\bf Declaration of LLM usage}
    \item[] Question: Does the paper describe the usage of LLMs if it is an important, original, or non-standard component of the core methods in this research? Note that if the LLM is used only for writing, editing, or formatting purposes and does \emph{not} impact the core methodology, scientific rigor, or originality of the research, declaration is not required.
    \item[] Answer: \answerYes{}
    \item[] Justification: LLMs are central to the proposed method as generators, refiners, and semantic oracles; the paper describes their role and the evaluated model configurations.

\end{enumerate}

\end{document}